\documentclass[fleqn,usenatbib,para,flushleft]{mnras}

\usepackage{newtxtext,newtxmath}
\usepackage{comment} 
\usepackage[T1]{fontenc}
\usepackage{subfigure}
\usepackage{adjustbox}
\usepackage{xcolor}
\usepackage{enumitem}

\DeclareRobustCommand{\VAN}[3]{#2}
\let\VANthebibliography\thebibliography
\def\thebibliography{\DeclareRobustCommand{\VAN}[3]{##3}\VANthebibliography}

\usepackage{graphicx}	
\usepackage{amsmath}	
\usepackage{epstopdf}
\usepackage{siunitx}    
\usepackage{booktabs}
\usepackage{longtable}
\usepackage{multicol}
\usepackage{lipsum}
\usepackage[labelfont=bf,labelsep=period]{caption}
\usepackage{threeparttablex}
\usepackage{pdflscape}

\usepackage{subcaption}

\DeclareGraphicsExtensions{.png,.pdf}
\usepackage{threeparttable}

\newcommand{\gaia}{\textsl{Gaia}\,}
\newcommand{\vpec}{\,$V_{\rm pec}$}
\newcommand{\be}{\,BeXRBs}
\newcommand{\sg}{\,SgXRBs}
\newcommand{\kms}{\,km\,s$^{-1}$}

\title[Natal kick segregation in HMXBs]{Further evidence for natal kick segregation by spectral type in high-mass X-ray binaries}

\author[P. Nuchvanichakul et al.]{Pornisara Nuchvanichakul,$^{1,2}$\thanks{E-mail: pornisara.nuchvanichakul@soton.ac.uk}
Poshak Gandhi,$^{2}$
Christian Knigge,$^{2}$
Yue Zhao,$^{2}$
\newauthor
Puji Irawati,$^{3}$
Suwicha Wanawichian$^{1,3}$
and Cordelia Dashwood Brown$^{2}$
\\
$^{1}$Department of Physics and Materials Science, Faculty of Science, Chiang Mai University, Chiang Mai, Thailand 50200\\
$^{2}$Department of Physics and Astronomy, University of Southampton, Highfield, Southampton SO17 1BJ, UK\\
$^{3}$National Astronomical Research Institute of Thailand, Chiang Mai, Thailand 50180
}

\date{Accepted XXX. Received YYY; in original form ZZZ}

\pubyear{20xx}

\begin{document}
\label{firstpage}
\pagerange{\pageref{firstpage}--\pageref{lastpage}}
\maketitle
\begin{abstract}
High-mass X-ray binaries (HMXBs) are systems in which a neutron star or black hole accretes material from a massive companion. HMXBs are expected to have experienced a supernova in their evolution. The impulsive kick associated with this event should affect the space velocity of the system in a way that depends on the nature and state of the progenitor binary. Here, we test whether the different evolutionary histories of HMXBs have left a detectable imprint on their peculiar velocities (\vpec). Using data from \gaia Data Release 3 (\gaia DR3), we first calculate the \vpec\ values for 63 well-known HMXBs hosting a black hole or neutron star and estimate the associated uncertainties via Monte Carlo re-sampling. We then analyse their distribution and check for differences between classes. Overall, \vpec\ estimates extend up to 100 \kms, but with Be/X-ray binaries (\be) favouring \vpec\ $\lesssim 40$\kms and supergiant X-ray binaries (\sg) favouring \vpec\ $\gtrsim 40$\kms. Based on a Kolmogorov–Smirnov (K-S) test, the null hypothesis that the peculiar velocities of both classes are drawn from the same parent distribution can be robustly rejected, irrespective of the background stellar velocity dispersion. Tests with binary population synthesis demonstrate that SgXRBs typically have shorter orbital periods and higher fractional mass loss than BeXRBs at supernova. We argue that the magnitude of \vpec\ could be used as a complementary feature to distinguish between Be and supergiant systems. These findings extend previous inferences based on two-dimensional kinematics from {\em Hipparcos}, and may be explained by the differing nature of the respective progenitors systems between the source classes at the instant of supernova.

\end{abstract}




\section{Introduction}
High-mass X-ray binaries (HMXBs) are mass-exchanging binary systems comprising a massive OB-star gravitationally bound to a compact object, either a neutron star or a black hole. The non-degenerate companions of HMXBs typically have masses in excess of $10\, M_{\sun}$. As a result, due to their immense brightness, HMXBs are excellent tools for probing Galactic star formation and compact objects. Estimating the total number of HMXB systems in the Milky Way is notoriously difficult due to unknown binary evolution physics. Over a hundred well-characterised HMXBs are currently known  \citep{liu_catalogue_2006,bird_ibis_2016, neumann_xrbcats_2023}.      
\\
 \indent Much still remains to be understood regarding the origin of HMXBs. During the supernova explosion, an impulsive kick can be imparted to the compact object. These kicks can have a decisive impact on the subsequent evolution and spatial distribution of HMXBs. Kicks can either be a result of symmetric mass loss (e.g., \citealp{blaauw_origin_1961}) or of asymmetric ejecta in the supernova explosion (e.g., \citealp{Chugai_1984,Dorofeev_1985,Arras_1999,janka_neutron_2017,Renzo_2019}). The strength of this ‘natal velocity kick’ ($V_{\rm nk}$) is thus a key ingredient for Galactic compact object population synthesis models and also for understanding the recent gravitational wave (GW) population (e.g. \citealp{dominik_double_2012}). Observationally, there are so far very few constraints on $V_{\rm nk}$, even for the bright HMXB population. One of the best-known cases is the BH-hosting HMXB Cyg X-1, which has been confirmed to have suffered only a very mild kick $V_{\rm nk}$ $\leq$ 20 \kms\ (e.g., \citealp{mirabel_formation_2017}). This has allowed the conclusive identification of the Cyg OB3 cluster as the natal site of this system \citep{rao_kinematic_2020}.

 \indent The low kick velocity inferred for Cyg X-1 is consistent with theoretical models in which kicks are momentum-conserving, i.e. in which $V_{\rm nk}$ scales inversely with black hole mass (at least qualitatively; e.g., \citealp{fryer_theoretical_2001}, \citealp{gandhi_gaia_2019}). However, this idea
 remains somewhat speculative (e.g., \citealp{repetto_galactic_2017}, \citealp{atri_potential_2019}, \citealp{gandhi_orbital_2020}). A recent study on distinct distribution between high-mass X-ray binaries (HMXBs) and low-mass X-ray binaries (LMXBs) provides empirical evidence of an inverse relationship between systemic velocity with respect to the total binary mass \citep{Zhao_2023}, albeit with much scatter.
 
 \indent HMXBs are typically amongst the brightest of the XRB population, as well as the youngest. These factors help to mitigate some of the key uncertainties related to kick inference and their evolutionary consequences, provided that key system parameters can be well constrained. HMXBs can be classified into three sub-classes, based on the nature of the secondary and the mass transfer process. Two of these are the OB-supergiant systems (hereafter \sg) and the Be/X-ray binaries (hereafter \be). In both of these sub-classes, the binary components are detached. In \sg\, the compact object accretes from the stellar wind of its massive companion; in \be\, accretion mainly takes place during  periastron passages, when the compact object passes through the decretion disk 
surrounding the rapidly rotating Be star \citep{fornasini}.The third sub-class is comprised of semi-detached systems, in which the non-degenerate companion loses mass to the compact object via Roche-lobe overflow (RLO) onto an accretion disc \citep{negueruela_stellar_2010}. 

Interestingly, previous studies with the {\em Hipparcos} mission \citep{chevalier_hipparcos_1998} found rather high transverse sky velocities for \sg $\, (V_{\rm t}$ $\sim$ 60 \kms and up to 90\,\kms), exceeding those of \be\, (average $V_{\rm t}$ value $\sim$ 11 \kms). 
The parameter $V_{\rm t}$ here is a two-dimensional (2-D) tracer of the three-dimensional (3-D) ‘peculiar velocity' (\vpec), with both referring to the motion relative to the Galactic rest frame (i.e., in excess of Galactic rotation, under the assumption that the system originated within the Galactic disc). Differences between the space velocities of different HMXB types could be suggestive of distinct evolutionary channels. However, before we can exploit this idea, we first need to confirm and quantify these differences. This is what we aim to undertake here.

\indent \gaia is a key mission of the European Space Agency's (ESA's) science programme, with its design goals relying heavily on astrometric, as well as photometric and spectroscopic surveys. The Third \gaia Data Release, known as \gaia DR3, published data from approximately 1.8 billion sources (\citeauthor{gaia_collaboration_gaia_2016} \citeyear{gaia_collaboration_gaia_2016}, \citeyear{gaia_collaboration_gaia_2022}). For bright and moderately faint sources, \gaia provides 5-parameter astrometry, including positions in right ascension ($\alpha$), declination ($\delta$), proper motions ($\mu_{\alpha}\cos\delta$, $\mu_{\delta}$), and parallaxes ($\pi$), with \texttt{G}-band magnitude ranging over $\sim$\,6 to 21 \citep{lindegren_gaia_2021}. Several studies have utilised \gaia data to investigate the kinematics and peculiar motions of X-ray binaries \citep[XRBs; cf.][]{gandhi_gaia_2019, atri_potential_2019, rao_kinematic_2020, Zhao_2023}. 
 
 The first focused study of HMXB sample kinematics was the aforementioned work by  \citet{chevalier_hipparcos_1998}, who found differences in the kinematics of Sg vs. BeXRBs, but were limited in having access to only {\em 2-D} (tangential) velocities from {\em Hipparcos} for a small ensemble of 17 systems. 
 Using updated \gaia\ Early Data Release 3 (EDR3) astrometry together with archival radial velocity information, \citet{fortin_constraints_2022} extracted 3-D kinematics of 35 {\em neutron-star} HMXBs, finding a \vpec\ distribution peaking around 116\,km\,s$^{-1}$. They also found a tendency for \sg\ to have higher \vpec\ values than \be\, but did not statistically quantify this trend.
 
 In this work, we build on previous studies by utilising the latest precise measurements of stellar kinematics provided by the \gaia\ DR3 catalogue \citep{gaia_collaboration_gaia_2022}. \gaia\ DR3 offers significant advancements over earlier releases, including updated radial velocities for approximately 34 million sources (compared to about 7 million in DR2/EDR3) and an extensive range of new data, such as detailed astrophysical parameters, quasar (QSO) candidates, solar system object data, and specific object studies \citep{gaia_collaboration_gaia_2022}. We include systemic radial velocities ($V_{\rm r}$) --- either direct measurements where available, or their estimates --- in our analysis, allowing us to compute the full 3-D \vpec\ values. We also include key black hole HMXB systems, enabling a more complete view of HMXB kinematics. Collating this sample allows us to test whether the evolutionary histories of the various HMXB classes leave an imprint on their kinematics. In Section~\ref{sec:gaia_counterpart_search}, we describe the \gaia\ counterpart search and sample selection process. Section~\ref{sec:distance} describes the \gaia\ distance estimation methodology. In Section~\ref{sec:vpeccomputation}, we describe the calculation of peculiar velocities. Section~\ref{sec:result} presents the results along with the statistical methods used in our analysis. In Section~\ref{sec:discussion}, we discuss our inferred peculiar velocities, the comparison of the two sub-classes, and completeness and selection effects. Finally, in Section~\ref{sec:summary}, we summarise our findings.

\section{Gaia counterpart search and association} \label{sec:gaia_counterpart_search}

Our starting parent sample comprises 114 systems selected from the fourth edition of the \citet{liu_catalogue_2006} HMXB catalogue, 8 additional HMXBs identified by \textit{INTEGRAL} \citep{bird_ibis_2016}, plus the XRB MWC 656, which has been suggested to host a BH \citep[][but see Section~\ref{sec:bhdiscussion}]{aleksic_magic_2015}. Additionally, we also included two rare cases of symbiotic X-ray binaries (SyXRBs): 4U 1954+31 and Swift J0850.8--4219 \citep{De_2023}. SyXRBs are characterised by the presence of a strongly magnetised neutron star and a late-type companion \citep{Bozzo2022}, and are included because they, like other sub-classes in HMXBs, have experienced a supernova event that formed the compact object. These two systems are the only two confirmed Galactic SyXRBs, and both have astrometric data available. Furthermore, we include Swift J0243.6+6124, \be\ recognised as the first Galactic ultraluminous X-ray pulsar (ULXP) \citep{Doroshenko2018,Tsygankov2018,Wilson2018}, which is not listed in the catalogues mentioned above.

A few sources were discarded from this parent sample. SAX J1819.3--2525, 1WGA J0648.0--4419 and IGR J12349--6434 are not HMXBs in the sense that we adopt in this work. The mass of the companion of SAX J1819.3--2525 is only 2.9 $M_{\sun}$, making this system an intermediate-mass X-ray binary (IMXB). Similarly, the companion of 1WGA J0648.0--4419 is a hot sub-dwarf \citep{jaschek_hd_1963}, and the optical counterpart of IGR J12349--6434 --- RT Cru --- has been classified as a symbiotic star which consists of a white dwarf (WD) and a red giant (RG) companion \citep{luna_nature_2007,Gromadzki2013AcA....63..405,ducci_rt_2016}. The systems OAO 1657--415, XTE J1543--568, and XTE J1858+034 have no known optical or infrared counterparts. AX J1749.2--2725 has an infrared counterpart, but no optical one and also lies close to a very bright, unrelated star \citep{karasev_axj17491-2733_2010}. On the other hand, we were able to retain two systems, 4U 1258--61 and IGR J16465--4507, that only have infrared counterparts with relatively large positional uncertainties (2\arcsec\ and 4\arcsec, respectively). We found close matches within 0.37\arcsec and 0.16\arcsec of their nominal positions, respectively, which we consider as genuine matches. This leaves 110 systems in our final parent sample.

We then queried the \gaia DR3 archive for sources within 5\arcsec\ of the literature positions. This search radius is larger than the positional uncertainty on most (but not all) HMXBs in our parent sample, so some of the queries yielded more than one possible \gaia\ counterpart. In all but two cases, the closest match was within 1\arcsec. We carried out checks to ensure the correct counterparts were identified among the \gaia sources within the search radius. In particular, the \gaia magnitudes and source identifiers were compared to the entries for the target HMXB in both \citet{liu_catalogue_2006} and relevant archived references in the \text{SIMBAD} database.\footnote{\url{http://simbad.u-strasbg.fr/simbad}} For the four HMXBs where the closest \gaia source was located more than 1\arcsec\ from the nominal position of the target, we carried out additional tests. These HMXBs are 4U 0115+634, 4U 0352+309, RX J0812.4--3114, and AX 1845.0--0433. For 4U 0115+634, we used the Robotic Optical Transient Experiment ($\emph{ROTSE}$) observations of its optical counterpart, V635 Cas, to confirm that the closest \gaia source is the correct counterpart \citep{Baykal_2005}. We also compared its position with the finding chart given by \citet{johnston_position_1978} and subsequently updated the position based on the optical observations of the HMXBs' companion reported by by \citet{Reig_2015}. We have chosen to adopt the positional coordinates provided by \citet{Reig_2015} as the reference in the literature. The optical counterpart of 4U 0352+309 is the bright variable star X~Per, which made it easy to ascertain that the nearest \gaia match is the correct counterpart. In the case of RX J0812.4--3114, we cross-check the finding chart and use the X-ray position provided by \citet{motch_new_1997}. For AX 1845.0–0433, a comparison between the finding chart provided by \citet{coe_discovery_1996} and the \gaia\ position confirms the accuracy of the source’s location. Consequently, we adopt the precise X-ray-based position reported by \citet{coe_discovery_1996}, which has a positional uncertainty of 0.5\arcsec, as the literature position.

All of the 110 \gaia counterparts have 5-parameter astrometric solutions in \gaia\ DR3. Distance estimation from parallax is not straightforward when the parallax uncertainty is large. To address this, we discard systems with fractional parallax errors larger than 20\% to avoid inaccurate and highly asymmetric error estimates. \citep[c.f.][]{bailer-jones_estimating_2015}. Our final HMXB sample is listed in Table~\ref{astrometric} and contains 63 systems, and their photometric and astrometric information is provided in Table~\ref{astrometric DR2}. These systems exhibited a range of \texttt{G}-band magnitudes between 6 and 14 and were predominantly located in the Galactic plane, with Galactic latitudes ranging from $b =$ \ang{-17.1} to $b =  $ \ang{5.7}.

\begin{table*}
    \centering
    \caption{Basic properties of the HMXB sample. For each source, we provide literature and \gaia DR3 coordinates, the offsets between these positions and the \gaia G-band magnitude (\texttt{G}).}
    \label{astrometric}
  
    \begin{adjustbox}{width=0.8\textwidth}
    \begin{tabular}{llccccccc} 
	\hline
		 & & &\multicolumn{2}{c}{Literature} &  \multicolumn{2}{c}{\gaia}  &  \\
   \cmidrule(lr){4-5}\cmidrule(lr){6-7}
		 No.& Source & Type & \multicolumn{1}{c}{$\alpha$ (J2000)} & \multicolumn{1}{c}{$\delta$ (J2000) }  & \multicolumn{1}{c}{$\alpha$ (J2015.5)} & \multicolumn{1}{c}{$\delta$ (J2015.5) }  &Offset& \texttt{G}  \\
		 & & &\multicolumn{1}{c}{(h:m:s)} & \multicolumn{1}{c}{(d:m:s)} &   \multicolumn{1}{c}{(h:m:s:)} & \multicolumn{1}{c}{(d:m:s)}  &(\arcsec)&mag\\
	
        \hline
	1&IGR J00370+6122 & Be & 00:37:10.00 & +61:21:35.0   &00:37:09.636 &+61:21:36.49 &3.03&9.45\\
	2&2S 0114+650     & Sg & 01:18:02.70&+65:17:30.0&01:18:02.694&+65:17:29.84&0.16&10.52\\
        3&4U 0115+634     & Be & 01:18:31.80&+63:44:33.0&01:18:31.966&+63:44:33.08&1.11&14.30\\
        4&IGR J01363+6610 & Be & 01:35:50.00&+66:12:40.0&01:35:49.852&+66:12:43.28&3.40&12.46\\
        5&RX J0146.9+6121 & Be & 01:47:00.20 & +61:21:23.7 &01:47:00.212& +61:21:23.66&0.08&11.22\\
	6&IGR J01583+6713 & Be & 01:58:18.20 & +67:13:25.9&01:58:18.491 & +67:13:23.46  &2.95&13.69\\
	7&1E 0236.6+6100  & Be & 02:40:31.70 & +61:13:46.0&02:40:31.660 &+61:13:45.59 &0.49&10.40\\
	8&V 0332+53       & Be & 03:34:59.90&+53:10:24.0&03:34:59.911&+53:10:23.30&0.70&14.20\\
        9&4U 0352+309     & Be & 03:55:23.10 & +31:02:45.0&03:55:23.080 &+31:02:45.01 &0.31&6.26\\
        10&XTE J0421+560  & Sg & 04:19:42.20&+55:59:59.0&04:19:42.135&+55:59:57.70&1.41&10.77\\
	11&RX J0440.9+4431& Be & 04:40:59.30 & +44:31:49.0&04:40:59.330 &+44:31:49.24  &0.40&10.40\\
	12&EXO 051910+3737.7 & Be &  05:22:35.20 & +37:40:34.0&05:22:35.230  &+37:40:33.58 &0.58&7.23\\
	13&1A 0535+262    & Be & 05:38:54.60 & +26:18:57.0 &05:38:54.570 &+26:18:56.79 &0.40&8.60\\
	14&1H 0556+286    & unclear & 05:55:55.10 & +28:47:06.0&05:55:55.040  &+28:47:06.39  &0.86&10.01\\
        15&IGR J06074+2205& Be & 06:07:26.60&+22:05:48.3&06:07:26.613&+22:05:47.75&0.58&12.17\\
        16&XTE J0658--073 & unclear & 06:58:17.30&--07:12:35.3&06:58:17.287&--07:12:35.18&0.22&11.99\\
        17&3A 0726--260   & unclear & 07 28 53.60&--26 06 29.0&07:28:53.578&--26:06:28.87&0.33&11.60\\
	18&1H 0739--529   & unclear & 07:47:23.60 & --53:19:57.0 &07:47:23.580 &--53:19:56.69 &0.37&7.54\\
        19&RX J0812.4--3114& unclear& 08:12:28.40&--31:14:51.0&08:12:28.356&--31:14:52.10&1.24&12.42\\
	20&4U 0900--40    & Sg & 09:02:06.90 & --40:33:17.0 &09:02:06.850 & --40:33:16.76  &0.58&6.74\\
        21&GRO J1008--57  & Be & 10:09:46.90&--58:17:35.5&10:09:46.955&--58:17:35.55&0.43&13.88\\
        22&RX J1037.5--5647& unclear & 10:37:35.20&--56:47:59.0&10:37:35.302&--56:47:55.82&3.29&11.24\\
	23&1A 1118--615   & Be & 11:20:57.20 & --61:55:00.0 &11:20:57.160 &--61:55:00.15 &0.31&11.59\\
        24&Cen X--3   & RLO& 11:21:15.10&--60:37:25.5&11:21:15.085&--60:37:25.59&0.14&12.88\\
        25&IGR J11215--5952& Sg & 11:21:46.81&--59:51:47.9&11:21:46.813&--59:51:47.93&0.03&9.77\\
	26&2S 1145--619   & Be & 11:48:00.00 & --62:12:25.0 &11:48:00.010 &--62:12:24.88 &0.13&8.65\\
        27&1E 1145.1--6141& Sg & 11:47:28.60&--61:57:14.0&11:47:28.546&--61:57:13.39&0.72&12.26\\
	28&4U 1223--624   & Sg & 12:26:37.60 & --62:46:13.0&12:26:37.550 & --62:46:13.29 &0.46&9.75\\
	29&1H 1249--637   & unclear & 12:42:50.30 & --63:03:31.0 &12:42:50.240 & --63:03:31.11 &0.44&5.14\\
	30&1H 1253--761   & unclear & 12:39:14.60 & --75:22:14.0&12:39:14.460 & --75:22:14.26&0.59&6.54\\
	31&1H 1255--567   & unclear & 12:54:36.90 & --57:10:07.0 &12:54:36.830 &--57:10:07.36 &0.69&5.15\\
	32&4U 1258--61    & Be & 13:01:17.10 & --61:36:07.0 &13:01:17.090 &--61:36:06.64 &0.37&12.65\\
        33&4U 1538--52    & Sg & 15:42:23.30&--52:23:10.0&15:42:23.352&--52:23:9.64&0.59&13.16\\
	34&1H 1555--552   & Be & 15:54:21.80 & --55:19:45.0 &15:54:21.760 & --55:19:44.36 &0.72&8.69\\
        35&IGR J16195--4945& Sg & 16:19:32.20&--49:44:30.7&16:19:32.183&--49:44:30.57&0.21&16.37\\
	36&IGR J16465--4507& Sg & 16:46:35.26 & --45:07:04.5 &16:46:35.260 & --45:07:04.66&0.16&13.48\\      
	37&4U 1700--37    & Sg & 17:03:56.80 & --37:50:39.0& 17:03:56.780  &--37:50:38.84 &0.33&6.42\\
        38&XTE J1739--302 & Sg & 17:39:11.58&--30:20:37.6&17:39:11.551&--30:20:37.73&0.40&12.64\\
	39&RX J1744.7--2713 & Be & 17:44:45.70 & --27:13:44.0 &17:44:45.760 &--27:13:44.51 &1.00&8.23\\
	40&IGR J17544--2619 & Sg & 17:54:25.28 & --26:19:52.6 &17:54:25.270 &--26:19:52.59 &0.11&11.66\\
	41&RX J1826.2--1450 & Be & 18:26:15.06 & --14:50:54.3 &18:26:15.060  &--14:50:54.37 &0.10&10.80\\
        42&AX 1845.0--0433  & Sg & 18:45:01.50&--04:33:55.5&18:45:1.589&--04:33:56.73&1.81&12.76\\
        43&3A 1909+048& Sg & 19:11:49.60&+04:58:58.0&19:11:49.562&+04:58:57.75&0.63&12.60 \\
	44&Cyg X--1 & Sg &  19:58:21.70 & +35:12:06.0& 19:58:21.670 & +35:12:05.69 &0.48&8.54\\
	45&RX J2030.5+4751 & Be &  20:30:30.80 & +47:51:51.0&20:30:30.840  &+47:51:50.65  &0.54&9.03\\
        46&GRO J2058+42& Be &  20:58:47.50&+41:46:37.0&20:58:47.534&+41:46:37.13&0.40&14.13\\
	47&SAX J2103.5+4545 & Be &   21:03:36.00 & +45:45:04.0 &21:03:35.700 &+45:45:05.52  &3.44&13.77\\
        48&1H 2138+579& Be &  21:39:30.60&+56:59:12.9&21:39:30.685&+56:59:10.39&2.61&13.82\\
	49&1H 2202+501 &  unclear &  22:01:38.20 & +50:10:05.0&22:01:38.210 & +50:10:04.63 &0.38&9.30\\
	50&4U 2206+543 &  Be &  22:07:56.20 & +54:31:06.0&22:07:56.230  &+54:31:06.36  &0.44&9.74\\
        51&SAX J2239.3+6116& Be &  22:39:20.90&+61:16:03.8&22:39:20.839&+61:16:26.59&0.49&14.10\\
        52& HD 259440&  Be &  06:32:59.26 & +05:48:01.2 & 06:32:59.257& +05:48:01.15 & 0.01 & 8.88 \\
	53& SAX J0635.2+0533  & Be &   06:35:18.28 & +05:33:06.3 & 06:35:18.279 & +05:33:06.28  & 0.01 & 12.50 \\
	54& IGR J08262--3736   &  Be &  08:26:13.65 & --37:37:11.9 & 08:26:13.651 & --37:37:11.82 & 0.06   & 12.16\\
	55& IGR J08408--4503   & Sg &   08:40:47.79 & --45:03:30.2 & 08:40:47.780 & --45:03:30.14  & 0.15   & 7.45   \\
	56& 2FGL J1019.0--5856 & Be &   10:18:55.59 & --58:56:46.0 & 10:18:55.574 & --58:56:45.94 & 0.11   & 12.27\\
	57& EXMS B1210--645    & Be &   12:13:14.79& --64:52:30.5 & 12:13:14.776  & --64:52:30.48 & 0.10   & 13.98 \\
	58& PSR B1259--63      & Be &   13:02:47.65 & --63:50:08.6 & 13:02:47.637 & --63:50:08.63  & 0.11   & 9.63\\
	59& IGR J21347+4737   & Be &   21:34:20.37 & +47:38:00.2 & 21:34:20.368& +47:38:00.16 & 0.05   & 14.00 \\
	60& MWC 656           & Be &   22:42:57.30 & +44:43:18.3 & 22:42:57.298 & +44:43:18.21 & 0.08   & 8.71\\
    61& SWIFT J0850.8--4219  & RSG & 08:50:40.08  & --42:11:52.3 & 08:50:40.081 & --42:11:51.45 & 0.92 & 13.35\\
    62& 4U 1954+31        & RSG & 19:55:42.27  & +32:05:48.8 & 19:55:42.336 & +32:05:48.95 &  0.82  & 8.36\\
    63& Swift J0243.6+6124 & Be & 02:43:40.33  & +61:26:02.8 & 02:43:40.424 & +61:26:03.76 & 1.17 & 12.39\\ 
        \hline
           
    \end{tabular}
    \end{adjustbox} 
\end{table*}

\begin{table*}
    \centering
    \caption{Astrometric measurements and radial velocities for the HMXB sample. For each source, we provide parallax, proper motions, and systemic radial velocity.}
    \label{astrometric DR2}
    \begin{adjustbox}{width=0.89\textwidth}
    \begin{threeparttable}
  
    \begin{tabular}{llllllll} 
	\hline
     
		 &  & &  Parallax &   \multicolumn{2}{c}{Proper motion}  & \multicolumn{2}{c}{Systemic radial velocity} \\
   \cmidrule(lr){5-6}\cmidrule(lr){7-8}
		 No.&Source& Spty & $\pi_{\rm t}$ & \multicolumn{1}{l}{$\mu_{\alpha}$cos$\delta$} & \multicolumn{1}{l}{$\mu_{\delta}$} & \multicolumn{1}{l}{$V_{\rm r}$} & \multicolumn{1}{l}{Ref.} \\
		& & (mas) & (mas y$^{-1}$)& (mas y$^{-1}$)& (\kms)&\\
		\hline
		1 & IGR J00370+6122& B0.5II--III  &   0.294$\pm$0.012 & --1.796$\pm$0.011 &--0.525$\pm$0.014 & --80.0$\pm$3.0&[1]\\
            2 & 2S 0114+650 & B0.5Ib    &   0.223$\pm$0.011&--1.243$\pm$0.009&0.761$\pm$0.012&--31.0$\pm$5.0&[2]\\
            3 & 4U 0115+634  & B0.2Ve   &   0.174$\pm$0.016&--1.684$\pm$0.013&0.504$\pm$0.017&--&--\\
            4 & IGR J01363+6610 & B1Ve&   0.174$\pm$0.011&--1.626$\pm$0.009&--0.027$\pm$0.011&--&--\\
		5 & RX J0146.9+6121 & B1Ve&   0.367$\pm$0.022 & --1.029$\pm$0.016 &--0.082$\pm$0.017 &--37.0$\pm$4.3 &[3]\\
		6 & IGR J01583+6713& Be &   0.167$\pm$0.013 & --1.198$\pm$0.011&0.300$\pm$0.013 &--& --\\
		7 & 1E 0236.6+6100 & B0Ve &   0.405$\pm$0.013 & --0.423$\pm$0.011&--0.256$\pm$0.012 & --41.41$\pm$0.60&[4]\\
            8 & V 0332+53   & O8.5Ve    &   0.180$\pm$0.020 &--0.268$\pm$0.020&0.440$\pm$0.020&--&--\\
		9 & 4U 0352+309& B0Ve&   1.668$\pm$0.037 & --1.282$\pm$0.053&--1.869$\pm$0.030 &1.0$\pm$0.9 &[5]\\
            10 & XTE J0421+560 & sgB[e] &   0.243$\pm$0.015 &--0.474$\pm$0.018&--0.510$\pm$0.013&--&--\\
		11 & RX J0440.9+4431 & B0.2Ve&  0.410$\pm$0.015 &0.101$\pm$0.016&--1.186$\pm$0.014 & --& --\\
		12 & EXO 051910+3737.7& B0IVpe &0.759$\pm$0.030& 1.305$\pm$0.041&--3.999$\pm$0.028 &-- &--\\
		13 & 1A 0535+262  & O9.7IIIe  &   0.560$\pm$0.023 &--0.590$\pm$0.031&--2.880$\pm$0.016 &--30.0$\pm$4.0&[6]\\
		14 & 1H 0556+286  & B5ne   &   0.626$\pm$0.030 & 0.634$\pm$0.034& --2.189$\pm$0.021 &34&[7]\\
            15 & IGR J06074+2205& Be &   0.166$\pm$0.018&0.573$\pm$0.020&--0.608$\pm$0.014&--&--\\
            16 &  XTE J0658--073& O9.7Ve &   0.174$\pm$0.015&--0.638$\pm$0.015&1.256$\pm$0.014&-- &--\\
            17 &  3A 0726--260 & O8--9Ve  &   0.130$\pm$0.017&--0.881$\pm$0.012&1.785$\pm$0.018&--&--\\
		18 & 1H 0739--529  & B7IV--Ve   &   1.544$\pm$0.021 & --4.572$\pm$0.027&8.530$\pm$0.028 &-- &--\\
            19 & RX J0812.4--3114& B0.2IVe&   0.149$\pm$0.012&--1.455$\pm$0.011&2.146$\pm$0.016&--&--\\
		20 & 4U 0900--40 & B0.5Ib    &   0.510$\pm$0.015 &--4.822$\pm$0.015&9.282$\pm$0.016 &--3.2$\pm$0.9 &[10]\\
            21 & GRO J1008--57 & B0e  &   0.282$\pm$0.013&--4.702$\pm$0.016&3.559$\pm$0.014&--&--\\
            22 & RX J1037.5--5647& B0V--IIIe&   0.197$\pm$0.016&--6.305$\pm$0.021&3.010$\pm$0.018&--&--\\
		23 & 1A 1118--615  & O9.5Ve  &   0.343$\pm$0.011 & --5.421$\pm$0.012&1.370$\pm$0.012 &--& --\\
            24 & Cen X--3 & O6.5II--II  &   0.145$\pm$0.014& --3.121$\pm$0.015&2.331$\pm$0.014&39$\pm$3&[11]\\
            25 & IGR J11215--5952& B1Ia &   0.138$\pm$0.012&--5.147$\pm$0.012&2.727$\pm$0.013&--&--\\
		26 & 2S 1145--619  & B0.2IIIe  &   0.489$\pm$0.017 & --6.226$\pm$0.017&1.598$\pm$0.018 &--&--\\
            27 & 1E 1145.1--6141& B2Iae &   0.121$\pm$0.010 & --6.226$\pm$0.010&2.362$\pm$0.012&--13.0$\pm$3.0&[12]\\
		28 & 4U 1223--624 & B1--1.5Ia   &   0.278$\pm$0.016 & --5.227$\pm$0.016&2.071$\pm$0.019&4.1$\pm$2.4&[13]\\
		29 & 1H 1249--637  & B0.5IIIe   &   2.299$\pm$0.077 &--12.857$\pm$0.070&--3.677$\pm$0.074 &22$\pm$7 &[14]\\
		30 & 1H 1253--761  & B7Vne  &   4.787$\pm$0.027 & --27.340$\pm$0.034&--8.934$\pm$0.040 &--20.0$\pm$7.4 &[14]\\
		31 & 1H 1255--567  & B5Ve  &   8.294$\pm$0.117&--28.386$\pm$0.088&--10.447$\pm$0.112&13.0$\pm$3.7 &[14]\\
		32 & 4U 1258--61   & B0.7Ve   &   0.542$\pm$0.014 & --4.341$\pm$0.012&--0.236$\pm$0.015&-- & --\\
            33 & 4U 1538--52   & B0Iab   &   0.176$\pm$0.015 &--6.711$\pm$0.015&--4.111$\pm$0.014&--158.0$\pm$11.0&[15]\\
		34 & 1H 1555--552  & B2IIIn  &   0.756$\pm$0.018 & --3.124$\pm$0.020&--3.223$\pm$0.016 &--& --\\
            35 & IGR J16195--4945& B1--2Ia &   0.391$\pm$0.051&--0.184$\pm$0.062&--0.545$\pm$0.044&--&--\\
		36 & IGR J16465--4507& B0.5I&   0.348$\pm$0.017&--1.759$\pm$0.022&--3.064$\pm$0.016&-- &--\\
		37 & 4U 1700--37   & O6.5Iaf+   &   0.668$\pm$0.026& 2.414$\pm$0.028&5.022$\pm$0.021&--60&[16]\\
            38 & XTE J1739--302 & O8Iab(f) &   0.534$\pm$0.048&--0.427$\pm$0.049&3.760$\pm$0.033&--&--\\
		39 & RX J1744.7--2713& B0.5V--IIIe&   0.822$\pm$0.024& --0.857$\pm$0.024&--2.296$\pm$0.016&--&--\\
		40 & IGR J17544--2619& O9Ib&   0.419$\pm$0.027 &--0.506$\pm$0.029&--0.668$\pm$0.018&--46.8$\pm$4.0 & [17]\\
		41 & RX J1826.2--1450 & ON6.5V((f))&  0.527$\pm$0.015 &7.425$\pm$0.014&--8.151$\pm$0.012&17.3$\pm$0.5&[18]\\
            42 & AX 1845.0--0433& O9.5I  &  0.184$\pm$0.024 &--1.366$\pm$0.024&--5.595$\pm$0.022&--&--\\
            43 & 3A 1909+048   & pec(BeBH)  &  0.135$\pm$0.023 &--3.027$\pm$0.024&--4.777$\pm$0.024&27$\pm$13&[11]\\
		44 & Cyg X--1 & O9.7Iab(BeBH)& 0.468$\pm$0.015 &--3.812$\pm$0.015&--6.310$\pm$0.017& --2.7$\pm$0.9,--7.0$\pm$0.5,--5.1$\pm$0.5 &[19], [20], [21]\\
		45 & RX J2030.5+4751 & B0.5V--IIIe&  0.437$\pm$0.016 & --2.714$\pm$0.020&--4.536$\pm$0.018&--  &--\\
            46 &  GRO J2058+42 & O9.5--B0IV--Ve   &  0.109$\pm$0.015 &--2.21$\pm$0.015&--3.351$\pm$0.017&--&--\\
		47 & SAX J2103.5+4545 & B0Ve&  0.161$\pm$0.013 & --3.505$\pm$0.014&--3.160$\pm$0.013&-- & --\\
            48 &  1H 2138+579  & B1--B2Ve   &  0.133$\pm$0.013&--2.964$\pm$0.014&--2.204$\pm$0.014&--&--\\
		49 & 1H 2202+501  & Be    &  0.896$\pm$0.013&2.365$\pm$0.015&--0.294$\pm$0.013&--16.8$\pm$ 2.5& [22] \\
		50 & 4U 2206+543  & O9.5Ve   &  0.320$\pm$0.014& --4.173$\pm$0.015&--3.317$\pm$0.014 &--62.7,--54.5$\pm$1.0 &[23], [24]\\
            51 &  SAX J2239.3+6116& B0V--B2IIIe& 0.136$\pm$0.014 &--2.344$\pm$0.015&--1.015$\pm$0.014&--&--\\
            52 & HD 259440   & B0pe    &  0.571$\pm$0.023 & --0.026$\pm$0.020 & --0.428$\pm$0.016 & 36.9$\pm$.8 &[25]\\
		53 & SAX J0635.2+0533 & B2V--B1IIIe & 0.159$\pm$0.015 & --0.419$\pm$0.013 & 0.405 $\pm$0.013 & --     & --\\
		54 & IGR J08262--3736  & OBV  & 0.194$\pm$0.010 & --2.367$\pm$0.009 & 3.177 $\pm$0.013 & -- & -- \\
		55 & IGR J08408--4503& O8.5Ib--II(f)p& 0.455$\pm$0.017 & --7.465$\pm$0.020 & 6.100 $\pm$0.019 & 15.3$\pm$0.5 & [26]\\
            56 & 2FGL J1019.0--5856& O6V & 0.232$\pm$0.010 & --6.454$\pm$0.013 & 2.256$\pm$0.013 & 33.0$\pm$3.0 & [27] \\
		57 & EXMS B1210--645  & B2V   & 0.301$\pm$0.018 & --5.953$\pm$0.016 & 0.450$\pm$0.021 & --42$\pm$11$^*$ & [28] \\
		58 & PSR B1259--63   & O9.5Ve   & 0.461$\pm$0.013 & --7.093$\pm$0.012 & --0.342$\pm$0.014 & 0.0$\pm$1.0  & [29] \\
		59 & IGR J21347+4737& B3V & 0.112$\pm$0.014 & --2.212$\pm$0.015 & --2.558$\pm$0.015 & -- & -- \\
		60 & MWC 656     & B3IVne+sh      & 0.509$\pm$0.018 & --3.478$\pm$0.016 & --3.159$\pm$0.017 &--14.1$\pm$2.1& [30] \\
        61   & SWIFT J0850.8--4219 & K3--5I& 0.132$\pm$0.014 & --3.533$\pm$0.015 &	4.217$\pm$0.016  & -- & -- \\
        62   & 4U 1954+31      & M4I        & 0.302$\pm$0.024 & --2.158$\pm$0.021 &	--6.071$\pm$0.026 & -- & -- \\
        63& Swift J0243.6+6124 & O9.5Ve & 0.192$\pm$0.011 & --0.729$\pm$0.010 & 0.134$\pm$0.012 & -- & -- \\
\hline
        
    \end{tabular}
    
        \begin{tablenotes}
            \footnotesize
            \item[] Spty: spectral type, $\pi_{\rm t}$: zeropoint-corrected parallax., ${}^*$: Value Case 2\\
            \item[] \textbf{References:} [1] \citet{grunhut_orbit_2014}; [2] \citet{koenigsberger_optical_2003}; [3] \citet{sarty_periodicities_2009}; [4] \citet{aragona_orbits_2009}; [5] \citet{grundstrom_joint_2007}; [6] \citet{hutchings_spectroscopic_1984}; [7] \citet{wilson_general_1953}; [8] \citet{stickland_spectroscopic_1994}; [9] \citet{thackeray_spectroscopic_1970}; [10] \citet{stickland_orbit_1997}; [11] \citet{duflot_vitesses_1995}; [12] \citet{hutchings_orbital_1987}; [13] \citet{kaper_vltuves_2006}; [14] \citet{kharchenko_astrophysical_2007}; [15] \citet{abubekerov_masses_2004}; [16] \citet{gies_binary_1986}; [17] \citet{nikolaeva_investigation_2013}; [18] \citet{casares_new_2011}; [19] \citet{gies_optical_1982}; [20] \citet{gies_wind_2003}; [21] \citet{gies_stellar_2008}; [22] \citet{chojnowski_high-resolution_2017}; [23] \citet{abt_stellar_1963}; [24] \citet{stoyanov_orbital_2014}; [25] \citet{moritani_orbital_2018}; [26] \citet{gamen_eccentric_2015}; [27] \citet{strader_optical_2015}; [28] \citet{Monageng2024}; [29] \citet{johnston_radio_1994}; [30] \citet{casares_be-type_2014}

        \end{tablenotes}

    \end{threeparttable}
    \end{adjustbox}    
\end{table*}

\section{Distance estimation with Gaia} \label{sec:distance}

Although \gaia provides precision astrometric measurements, these are not free of systematic biases \citep{lindegren_basic_2010}. One of the biases impacting the reported parallaxes manifests as a zero-point offset whose size depends on the magnitude, colour, and ecliptic latitude of the source \citep{lindegren_gaia_2021}. In order to estimate the parallax zero-point offset values (ZP) for our sources, we used the Python package {\sc gaiadr{\scriptsize 3}\_zeropoint}\footnote{\url{https://gitlab.com/icc-ub/public/gaiadr3_zeropoint}}, which implements the corrections described in \citet{lindegren_gaia_2021}. Following \citet{groenewegen_parallax_2021}, we then obtain an estimate of the true parallax ($\pi_{\rm t}$) by applying a correction as $\pi_{\rm t} = \pi_{\rm o} - {\rm ZP}$, where $\pi_{\rm o}$ is the observed parallax from \gaia DR3.

We then use the zero-point-corrected parallaxes to estimate distances. Since all sources in our sample have parallax uncertainties less than 20\%, parallax inversion, i.e. $r_{\gaia} = 1/\pi_{\rm t}$, is a fair estimator of their distances \citep{bailer-jones_estimating_2015}.

\begin{landscape}
\begin{table}
\centering
\caption{Inferred sample distances and peculiar velocities, together with other source properties from the literature.}
\label{d}
\begin{threeparttable}
\begin{tabular}{lllllllllllllll}
\hline
    &        &       \multicolumn{2}{c}{Pulse Period} & \multicolumn{2}{c}{Orbital Period} & $V_{\rm pec}$ & $r_{\gaia}$ & \multicolumn{2}{c}{$r_{\rm lit}$} & \multicolumn{2}{c}{$M_{\rm 1}$} & \multicolumn{2}{c}{$M_{\rm 2}$} \\
    \cmidrule(lr){3-4}\cmidrule(lr){5-6}\cmidrule(lr){9-10} \cmidrule(lr){11-12} \cmidrule(lr){13-14}
No. & Source  & \multicolumn{1}{l}{(s)} & \multicolumn{1}{l}{Ref.}         & \multicolumn{1}{l}{(day)} & \multicolumn{1}{l}{Ref.}      & (\kms)          & (kpc)       & \multicolumn{1}{l}{(kpc)} & \multicolumn{1}{l}{Ref.} & \multicolumn{1}{l}{(M$_\odot$)} & \multicolumn{1}{l}{Ref.}& \multicolumn{1}{l}{(M$_\odot$)} & \multicolumn{1}{l}{Ref.} \\
\hline
1 & IGR J00370+6122 & 346$\pm$6 & [1] & 15.665$\pm$0.006 & [2] & 33.0$_{-3.4}^{+3.4}$  & 3.40$\pm$0.14 & 3.3 & [3] &1.4$^\dagger$ & -- & 10.0$_{-5.0}^{+5.0}$ & [4] \\
2 & 2S 0114+650  & 9475$\pm$25 & [5] & 11.6 & [6] & 36.2$_{-4.2}^{+5.1}$ & 4.48$\pm$0.23 & 7.0$\pm$3.6 & [7] &1.4$^\dagger$ & -- & 16.0$_{-2.0}^{+2.0}$ & [8]\\
3 & 4U 0115+634  & 3.61 & [9] & 24.309$\pm$0.021 & [9] & 19.7$_{-1.6}^{+1.8}*$ & 5.74$\pm$0.54 & 7--8 & [10]&--&--&--&-- \\
4 & IGR J01363+6610  & -- & -- & 159$\pm$2 & [11] & $6.9_{-1.6}^{+1.8}*$ & 5.75$\pm$0.36 & 2 & [3]&--&--&--&-- \\
5 & RX J0146.9+6121  & 1404.2$\pm$1.2 & [12] & 330 & [13] & 9.2$_{-1.4}^{+2.1}$ & 2.73$\pm$0.16 & 2.2, 2.5$\pm$0.6 & [14], [15]&1.4$^\dagger$ & -- & 11.0$_{-2.0}^{+2.0}$ & [16] \\
6 & IGR J01583+6713  & 469.2 & [17] & 3--12 & [18] & 7.3$_{-1.7}^{+2.1}*$ & 5.99$\pm$0.47 & 3.4$\pm$0.8, 4.0$\pm$0.4, 6.4 & [15], [17], [19]&--&--&--&-- \\
7 & 1E 0236.6+6100  & -- & -- & 26.4960$\pm$0.0028 & [20] & 11.0$_{-0.9}^{+0.8}$ & 2.47$\pm$0.08 & 2.0, 2.0$\pm$0.2 & [21], [22]&1.4$^\dagger$ & -- & 12.5$_{-2.5}^{+2.5}$ & [23]\\
8 & V 0332+53  & 4.4 & [24] & 34.25$\pm$0.10 & [24] & 19.3$_{-1.6}^{+1.8}*$ & 5.57$\pm$0.63 & 7 & [25]&--&--&--&-- \\
9 & 4U 0352+309  & 835 & [26] & 250$\pm$0.6 & [27] & 12.2$_{-0.7}^{+0.9}$ & 0.60$\pm$0.01 & 1.30$\pm$0.40, 0.7$\pm$0.3 & [28], [29]&1.4$^\dagger$ & -- & 14.0$_{-3.0}^{+3.0}$ & [30]\\
10 & XTE J0421+560  & -- & -- & 19.41$\pm$0.02 & [31] & 12.3$_{-1.7}^{+1.9}*$ & 4.11$\pm$0.25 & >5 & [32]&--&--&--&--  \\
11 & RX J0440.9+4431  & 202.5$\pm$0.5 & [33] & 150$\pm$0.2 & [34] & 3.4$_{-1.6}^{+1.5}*$ & 2.44$\pm$0.09 & 3.3$\pm$0.50 & [35] &--&--&--&-- \\
12 & EXO 051910+3737.7 & -- & -- & -- & -- & 16.7$_{-1.6}^{+1.7}*$ & 1.32$\pm$0.05 & 1.7$\pm$0.1 & [36] &--&--&--&-- \\
13 & 1A 0535+262  & 104 & [37] & 111$\pm$0.4 & [24], [37] & 41.7$_{-3.8}^{+4.2}$ & 1.79$\pm$0.07 & 2.00$\pm$0.70, 2.9 & [21], [38] & 1.6$_{-0.6}^{+0.6}$ & [39] &7.5$_{-2.5}^{+2.5}$ & [39] \\
14 & 1H 0556+286 & -- & -- & -- & -- & 25.8$_{-0.8}^{+1.0}$ & 1.60$\pm$0.08 & 0.83 & [40]&--&--&--&-- \\
15 & IGR J06074+2205 & 373.2 & [41] & -- & -- & 25.5$_{-2.6}^{+3.0}*$ & 6.03$\pm$0.66 & 4.4$\pm$1.0 & [42] &--&--&--&--\\
16 & XTE J0658--073  & 160.4$\pm$0.4 & [43] & 101.20 & [43] & 12.4$_{-1.8}^{+2.3}*$ & 5.74$\pm$0.50 & 3.9$\pm$0.1 & [43]&--&--&--&-- \\
17 & 3A 0726--260  & 103.2 & [44] & 34.5 & [44] & 15.3$_{-3.0}^{+5.6}*$ & 7.67$\pm$1.00 & 4.6$\pm$1.3, 6.1$\pm$0.3 & [45], [46]&--&--&--&-- \\
18 & 1H 0739--529 & -- & -- & -- & -- & 8.2$_{-1.5}^{+1.5}*$ & 0.65$\pm$0.01 & 0.52 & [47]&--&--&--&-- \\
19 & RX J0812.4--3114  & 31.8851 & [48] & 81.3 & [49] & 28.5$_{-3.2}^{+3.0}*$ & 6.70$\pm$0.55 & 11.4, 8.8$\pm$4.0 & [50], [51]&--&--&--&-- \\
20 & 4U 0900--40  & 283 & [26] & 8.97 & [52] & 58.0$_{-1.8}^{+2.2}$ & 1.96$\pm$0.06 & 1.90$\pm$0.20 & [53] & 2.1$_{-0.2}^{+0.2}$ & [54] &26.0$_{-1.0}^{+1.0}$ & [54]\\
21 & GRO J1008--57  & 93.587$\pm$0.005 & [55] & 247.5 & [56] & 19.9$_{-1.8}^{+2.0}*$ & 3.54$\pm$0.17 & 5 & [57]&--&--&--&-- \\
22 & RX J1037.5--5647  & 860$\pm$2 & [33] & 61.0$\pm$0.2 & [58] & 21.7$_{-4.1}^{+5.9}*$ & 5.06$\pm$0.42 & 5 & [50] &--&--&--&--\\
23 & 1A 1118--615  & 405 & [26] & 24 & [59] & 21.5$_{-1.9}^{+2.3}*$ & 2.92$\pm$0.10 & 5$\pm$2 & [60]&--&--&--&-- \\
24 & Cen X--3  & 4.84 & [26] & 2.09 & [61] & 96.5$_{-3.5}^{+3.5}$ & 6.89$\pm$0.65 & 10$\pm$1 & [62] & 1.6$_{-0.1}^{+0.1}$ & [54] &24.0$_{-1.0}^{+1.0}$ & [54]\\
25 & IGR J11215--5952 & 186.78$\pm$0.3 & [63] & 165 & [64] & 49.1$_{-2.1}^{+2.6}*$ & 7.27$\pm$0.65 & 8, 6.2 & [65], [66]&--&--&--&-- \\
26 & 2S 1145--619   & 292.4 & [26] & 187.5 & [26] & 11.5$_{-1.5}^{+1.5}*$ & 2.05$\pm$0.07 & 3.1, 1.4 & [67], [68] &1.4$^\dagger$ & -- & 13.0$_{-2.0}^{+2.0}$ & [69]\\
27 & 1E 1145.1--6141  & 298$\pm$4 & [70] & 14.37$\pm$0.02 & [71] & 55.3$_{-8.2}^{+14.0}$ & 8.29$\pm$0.70 & 8 & [72]& 1.7$_{-0.3}^{+0.3}$ & [73] &14.0$_{-4.0}^{+4.0}$ & [73] \\
28 & 4U 1223--624  & 696 & [74] & 41.498$\pm$0.002 & [75] & 54.3$_{-3.5}^{+3.6}$ & 3.60$\pm$0.21 & 5.3, 1.8$\pm$0.4, 4.1 & [76], [77], [78] & 1.9$_{-0.6}^{+0.6}$ & [75] &43.0$_{-10.0}^{+10.0}$ & [75]\\
29 & 1H 1249--637 & 14200 & [79] & -- & -- & 23.8$_{-7.3}^{+7.1}$ & 0.44$\pm$0.01 & 0.30$_{-0.06}^{+0.04}$ & [47] &1.4$^\dagger$ & -- & 9.6 & [80]\\
30 & 1H 1253--761  & -- & -- & -- & -- & 28.2$_{-6.5}^{+7.7}$ & 0.21$\pm$1e-3 & 0.236$_{-0.039}^{+0.029}$ & [47] &1.4$^\dagger$ & -- & 7.5$^\dagger$ & [81]\\
31 & 1H 1255--567  & -- & -- & -- & -- & 10.8$_{-3.6}^{+4.1}$ & 0.12$\pm$2e-3 & 0.11$_{-0.008}^{+0.007}$ & [47] & -- & -- & -- & -- \\
32 & 4U 1258--61 & 272 & [82] & 133 & [26] & 24.8$_{-1.6}^{+1.5}*$ & 1.85$\pm$0.05 & 2.4$\pm$0.5 & [83] & -- & -- & -- & -- \\
33 & 4U 1538--52 & 529 & [84], [85] & 3.73 & [84], [85] & 91.7$_{-10.3}^{+10.8}$ & 5.69$\pm$0.50 & 6.4$\pm$1.0, 4.5 & [86], [87] & 1.0$_{-0.2}^{+0.2}$ & [54] &16.0$_{-2.0}^{+2.0}$ & [54] \\
34 & 1H 1555--552  & -- & -- & -- & -- & 6.3$_{-1.5}^{+1.5}*$ & 1.32$\pm$0.03 & 0.96 & [88]&1.4$^\dagger$ & -- & 19.4$_{-5.0}^{+5.0}$ & [89] \\
35 & IGR J16195--4945 & -- & -- & 3.945$\pm$0.005 & [90] & 41.1$_{-2.5}^{+5.2}*$ & 2.56$\pm$0.33 & 7 & [91] & -- & -- & -- & --\\
36 & IGR J16465--4507  & 228$\pm$6 & [92] & 30.32$\pm$0.02 & [93] & 21.0$_{-1.6}^{+1.7}*$ & 2.88$\pm$0.14 & 12.50, 9.50$_{-5.7}^{+14.1}$ & [93], [94] & -- & -- & -- & --\\
37 & 4U 1700--37 & 67.4? & [95] & 3.41 & [96] & 72.7$_{-5.8}^{+6.4}$ & 1.50$\pm$0.06 & 1.90$\pm$0.30 & [97] & 2.0$_{-0.2}^{+0.2}$ & [54] &46.0$_{-5.0}^{+5.0}$ & [54] \\
38 & XTE J1739--302  & -- & -- & 51.47$\pm$0.02 & [98] & 62.5$_{-2.7}^{+4.1}*$ & 1.87$\pm$0.17 & 2.3 & [99] & -- & -- & -- & --\\
39 & RX J1744.7--2713  & -- & -- & -- & -- & 6.6$_{-1.5}^{+1.6}*$ & 1.22$\pm$0.04 & 0.80 & [50]& -- & -- & -- & -- \\
40 & IGR J17544--2619  & 11.58$\pm$0.03 & [100] & 12.172$\pm$0.007 & [101] & 45.1$_{-4.2}^{+3.9}$ & 2.39$\pm$0.15 & 8, 2-4 & [102], [103] &1.4$^\dagger$ & -- & 23.0$_{-2.0}^{+2.0}$ & [104]\\
41 & RX J1826.2--1450  & -- & -- & 3.91 & [105] & 90.3$_{-2.9}^{+3.0}$ & 1.90$\pm$0.05 & 2.5$\pm$0.1 & [105] & 3.7$_{-1.0}^{+1.3}$ & [105] &22.9$_{-2.9}^{+3.4}$ & [105]\\
42 & AX 1845.0--0433  & -- & -- & 5.72 & [106] & 54.5$_{-2.5}^{+8.0}*$ & 5.44$\pm$0.70 & 3.6, 6.4$\pm$0.76 & [107], [108]& -- & -- & -- & -- \\
43 & 3A 1909+048  & -- & -- & 13.1 & [109] & 57.0$_{-9.4}^{+10.2}$ & 7.42$\pm$1.28 & 5.5$\pm$0.2 & [110] & 4.3$_{-0.4}^{+0.4}$ & [111] &11.3$_{-0.6}^{+0.6}$ & [111]\\
44 & Cyg X--1  & -- & -- & 5.60 & [112] & 20.8$_{-1.4}^{+1.4}$ & 2.14$\pm$0.07 & 1.86$\pm$0.12 & [113] & 21.2$_{-2.2}^{+2.2}$ & [114] &40.6$_{-7.1}^{+7.7}$ & [114]\\

\hline
\end{tabular}
\end{threeparttable}
\end{table}
\end{landscape}

\begin{landscape}
\begin{table}
\centering
\contcaption{}
\begin{threeparttable}
\begin{tabular}{lllllllllllllll}
\hline
    &        &       \multicolumn{2}{c}{Pulse Period} & \multicolumn{2}{c}{Orbital Period} & $V_{\rm pec}$ & $r_{\gaia}$ & \multicolumn{2}{c}{$r_{\rm lit}$} & \multicolumn{2}{c}{$M_{\rm 1}$} & \multicolumn{2}{c}{$M_{\rm 2}$} \\
    \cmidrule(lr){3-4}\cmidrule(lr){5-6}\cmidrule(lr){9-10} \cmidrule(lr){11-12} \cmidrule(lr){13-14}
No. & Source  & \multicolumn{1}{l}{(s)} & \multicolumn{1}{l}{Ref.}         & \multicolumn{1}{l}{(day)} & \multicolumn{1}{l}{Ref.}      & (\kms)          & (kpc)       & \multicolumn{1}{l}{(kpc)} & \multicolumn{1}{l}{Ref.} & \multicolumn{1}{l}{(M$_\odot$)} & \multicolumn{1}{l}{Ref.}& \multicolumn{1}{l}{(M$_\odot$)} & \multicolumn{1}{l}{Ref.} \\
\hline

45 & RX J2030.5+4751  & -- & -- & -- & -- & 1.7$_{-1.5}^{+1.5}*$ & 2.29$\pm$0.08 & 2.20 & [50] & -- & -- & -- & --\\
46 & GRO J2058+42 & 198 & [115] & 55.03 & [115] & 19.9$_{-3.0}^{+9.8}*$ & 9.18$\pm$1.28 & 9$\pm$1 & [3]&1.4$^\dagger$ & -- & 18.0$^\dagger$ & [80] \\
47 & SAX J2103.5+4545 & 358.62 & [116] & 12.68$\pm$0.25 & [116] & 30.2$_{-1.7}^{+2.2}*$ & 6.20$\pm$0.50 & 6.50$\pm$0.90 & [117]& -- & -- & -- & -- \\
48 & 1H 2138+579 & 66.2 & [26], [118] & 20.85 & [119] & 20.3$_{-2.4}^{+4.2}*$ & 7.50$\pm$0.71 & 3.8$\pm$0.6 & [120] & -- & -- & -- & --\\
49 & 1H 2202+501 & -- & -- & -- & -- & 30.3$_{-1.1}^{+1.3}$ & 1.12$\pm$0.02 & 0.70 & [47] & -- & -- & -- & --\\
50 & 4U 2206+543  & 5554$\pm$9 & [121] & 19.25$\pm$0.8 & [122] & 24.1$_{-1.6}^{+1.8}$ & 3.12$\pm$0.13 & 2.6 & [123] &1.4$^\dagger$ & -- & 23.5$_{-8.0}^{+14.5}$ & [124]\\
51 & SAX J2239.3+6116  & 1247 & [125] & 262 & [126] & 21.6$_{-2.3}^{+2.9}*$ & 7.36$\pm$0.76 & 4.4 & [126]& -- & -- & -- & -- \\
52 & HD 259440  & -- & -- & 308.0$\pm$26.0 & [127] & 9.4$_{-1.0}^{+1.3}$ & 1.75$\pm$0.07 & 11--17 & [128] &1.4& [128] & 15.7$_{-2.5}^{+2.5}$ & [128]\\
53 & SAX J0635.2+0533  & 0.034 & [129] & 11.2$\pm$0.5 & [130] & 10.5$_{-1.5}^{+1.6}*$ & 6.29$\pm$0.59 & 2.5--5 & [131] & -- & -- & -- & --\\
54 & IGR J08262--3736 & -- & -- & -- & -- & 11.8$_{-2.1}^{+2.4}*$ & 5.15$\pm$0.26 & 6.1 & [132] & -- & -- & -- & --\\
55 & IGR J08408--4503  & -- & -- & 9.54 & [133] & 41.0$_{-2.6}^{+2.2}$ & 2.20$\pm$0.08 & 2.7 & [134]&1.4$^\dagger$ & -- & 33.0$^\dagger$ & [135]  \\
56 & 2FGL J1019.0--5856  & -- & -- & 16.54 & [136] & 35.2$_{-2.2}^{+2.0}$ & 4.31$\pm$0.19 & 6.4$_{-0.7}^{+1.7}$ & [137] &1.4$^\dagger$ & [138] & 23.0$_{-3.0}^{+3.0}$ & [138]\\
57 & EXMS B1210--645 & -- & -- & 6.7 & [139] & 27.4$_{-11.0}^{+9.5}$ & 3.33$\pm$0.20 & 2.8 & [140]& -- & -- & -- & -- \\
58 & PSR B1259--63  & 0.0478 & [141] & 1236.72 & [142] & 24.1$_{-1.4}^{+1.3}$ & 2.17$\pm$0.06 & 2.6$_{-0.3}^{+0.4}$ & [142] & 1.4$^\dagger$ & -- & 22.5$_{-7.5}^{+7.5}$ & [142] \\
59 & IGR J21347+4737 & -- & -- & -- & -- & 18.1$_{-3.9}^{+9.8}*$ & 8.93$\pm$1.13 & 5.8 & [140] & -- & -- & -- & --\\
60 & MWC 656  & -- & -- & 60.37$\pm$0.04 & [143] & 10.8$_{-2.0}^{+2.1}$ & 1.97$\pm$0.07 & 2.6$\pm$1.0 & [144] & 4.1$_{-1.4}^{+1.4}$ & [144]& 7.8$_{-2.0}^{+2.0}$ & [143]\\
61 & SWIFT J0850.8--4219  & -- & -- & -- & -- & 68.8${}^{ +18.9}_{-10.0}\ast$ & 7.57$\pm$0.80 & 12 & [145]& -- & -- & -- & -- \\
62 & 4U 1954+31  & 19400 & [146] & 1296.64 & [146] & 19.7$_{-2.1}^{+2.2}*$ & 3.31$\pm$0.27 & 3.4$_{-0.2}^{+0.3}$ & [147] &1.4$^\dagger$ & -- & 9.0$_{-4.0}^{+4.0}$ & [146]\\
63& Swift J0243.6+6124 & 9.86 & [148], [149], [150] & 28.3$\pm$0.2 & [151] & 10.4$_{-1.5}^{+1.6}*$ & 5.20$\pm$0.31 & 4.5$\pm$0.5 & [152] & -- & -- & -- & -- \\
\hline
\end{tabular}
\begin{tablenotes}  
\footnotesize
\item[] \footnotesize{${}^{\ast}$: $V_{\rm pec,3D}^{\rm iso}$, ${}^{\dagger}$: values for which no constraints are available in the literature.}\\
\item[] \footnotesize{\textbf{References:} [1] \citet{inZand2007}; [2] \citet{denHartog2004}; [3] \citet{reig_identification_2005}; [4] \citet{gonzalez-galan_astrophysical_2014}; [5] \citet{Wang_2011}; [6] \citet{crampton_supergiant_1985}; [7] \citet{reig_astrophysical_1996}; [8] \citet{Hu2017}; [9] \citet{rappaport_orbital_1978}; [10] \citet{Negueruela_Okazaki_2001}; [11] \citet{Corbet_2010ATel}; [12] \citet{Haberl_1998}; [13] \citet{sarty_periodicities_2009}; [14] \citet{coe_infrared_1993}; [15] \citet{reig_long-term_2016}; [16] \citet{reig1997}; [17] \citet{kaur_multiwavelength_2008}; [18] \citet{Wang_2010}; [19] \citet{masetti_unveiling_2006}; [20] \citet{Gregory_2002ApJ...575..427G}; [21] \citet{steele_distances_1998}; [22] \citet{Frail_1991}; [23] \citet{casares_orbital_2005}; [24] \citet{Stella_1985}; [25] \citet{negueruela_bex-ray_1999}; [26] \citet{Nagase_1989}; [27] \citet{delgado-marti_orbit_2001}; [28] \citet{fabregat_astrophysical_1992}; [29] \citet{lyubimkov_fundamental_1997}; [30] \citet{grundstrom_joint_2007}; [31] \citet{Barsukova_2005ATel}; [32] \citet{robinson_high-dispersion_2002}; [33] \citet{Reig_1999}; [34] \citet{Ferrigno_2013}; [35] \citet{reig_long-term_2005}; [36] \citet{polcaro_search_1989}; [37] \citet{Priedhorsk_1983}; [38] \citet{lyuty_one_2000}; [39] \citet{hutchings_spectroscopic_1984}; [40] \citet{bonnet-bidaud_x-ray_1981}; [41] \citet{Reig_Zezas_2018}; [42] \citet{reig_optical_2010}; [43] \citet{mcbride_study_2006}; [44] \citet{corbet_rxte_1997}; [45] \citet{corbet_optical_1984}; [46] \citet{negueruela_optical_1996}; [47] \citet{chevalier_hipparcos_1998}; [48] \citet{Reig_Peele_1999}; [49] \citet{Corbet_Peele_2000}; [50] \citet{motch_new_1997}; [51] \citet{reig_bex-ray_2001}; [52] \citet{van_der_Klis_1984}; [53] \citet{sadakane_ultraviolet_1985}; [54] \citet{Falanga2015}; [55] \citet{Stollber_1993}; [56] \citet{Okazaki_2001}; [57] \citet{coe_discovery_1994}; [58] \citet{Cusumano_2013}; [59] \citet{Staubert_2011}; [60] \citet{janot-pacheco_photometric_1981}; [61] \citet{Kelley_1983}; [62] \citet{hutchings_centaurus_1979}; [63] \citet{Swank_2007}; [64] \citet{Sidoli_2007}; [65] \citet{negueruela_hd_2005}; [66] \citet{sidoli_igr_2006}; [67] \citet{negueruela_nature_1998}; [68] \citet{hutchings_x-ray_1981}; [69] \citet{Stevens1997}; [70] \citet{Lamb_1980}; [71] \citet{Ray_2002}; [72] \citet{ilovaisky_nature_1982}; [73] \citet{hutchings_orbital_1987}; [74] \citet{Sato_1986}; [75] \citet{kaper_vltuves_2006}; [76] \citet{kaper_wray_1995}; [77] \citet{parkes_spectral_1980}; [78] \citet{leahy_rxteasm_2002}; [79] \citet{Torrej_2001}; [80] \citet{Zorec2005}; [81] \citet{Waters1989}; [82] \citet{Priedhorsky_1983}; [83] \citet{parkes_shell_1980}; [84] \citet{Becker_1977}; [85] \citet{Davison_1977}; [86] \citet{reynolds_optical_1992}; [87] \citet{clark_chandraobservations_2004}; [88] \citet{grillo_einstein_1992}; [89] \citet{Fairlamb2015}; [90] \citet{Cusumano_2016}; [91] \citet{sidoli_soft_2005}; [92] \citet{Lutovinov_2005}; [93] \citet{clark_orbital_2010}; [94] \citet{smith_circumstantial_2004}; [95] \citet{Murakami_1984}; [96] \citet{Jones_1973}; [97] \citet{ankay_origin_2001}; [98] \citet{Drave_2010}; [99] \citet{negueruela_optical_2006}; [100] \citet{Romano_2015}; [101] \citet{nikolaeva_investigation_2013}; [102] \citet{gonzalez-riestra_xmm-newton_2004}; [103] \citet{pellizza_igr_2006}; [104] \citet{Bikmaev2017}; [105] \citet{casares_possible_2005}; [106] \citet{Goossens_2013}; [107] \citet{coe_discovery_1996}; [108] \citet{Coleiro_2013}; [109] \citet{crampton_ss_1981}; [110] \citet{blundell_symmetry_2004}; [111] \citet{Picchi2020}; [112] \citet{LaSala_1998}; [113] \citet{Reid_Trigonometric_2011}; [114] \citet{miller-jones_cygnus_2021}; [115] \citet{Wilson_1998}; [116] \citet{Baykal_2000}; [117] \citet{reig_discovery_2004}; [118] \citet{Koyama_1991}; [119] \citet{McBride_2007}; [120] \citet{bonnet-bidaud_identification_1998}; [121] \citet{Finger_2010}; [122] \citet{Corbet_2206_2007}; [123] \citet{blay_multiwavelength_2006}; [124] \citet{Hambaryan2022}; [125] \citet{intZand_2001}; [126] \citet{int_zand_transient_2000}; [127] \citet{Moritani_2018}; [128] \citet{aragona_hd_2010}; [129] \citet{Cusumano_2000}; [130] \citet{Kaaret_2000}; [131] \citet{mereghetti_low_2009}; [132] \citet{Masetti_2012}; [133] \citet{Gamen_2015}; [134] \citet{leyder_hard_2007}; [135] \citet{gamen_eccentric_2015}; [136] \citet{An_2015}; [137] \citet{marcote_refining_2018}; [138] \citet{Strader2015}; [139] \citet{walter_high-mass_2015}; [140] \citet{masetti_unveiling_2009}; [141] \citet{Manchester_1995}; [142] \citet{miller-jones_geometric_2018}; [143] \citet{williams_be_2010}; [144] \citet{casares_binary_2012}; [145] \citet{De_2023}; [146] \citet{Hinkle2020}; [147] \citet{Fortin_cat_2023}; [148] \citet{Kennea2017ATel}; [149] \citet{JenkeWilson2017ATel}; [150] \citet{Bahramian2017ATel}; [151] \citet{Doroshenko2018}; [152] \citet{J0243Reig2020}}
\end{tablenotes}
\end{threeparttable}
\end{table}
\end{landscape}

\section{Kinematics}
\label{sec:vpeccomputation}

All HMXBs in our sample have five-parameter astrometric solutions in DR3, viz. celestial positions (right ascension $\alpha$ and declination $\delta$), proper motions ($\mu_{\alpha}\cos\delta$, hereafter $\mu_{\alpha*}$, and $\mu_{\delta}$) and parallax ($\pi$). Note that we use the estimate of the true parallax ($\pi_{\rm t}$) defined  
in Section~\ref{sec:distance} for estimating peculiar velocities (\vpec). 

In order to estimate the 3-D \vpec\, of a source, we need one final measurement --- the systemic radial velocity ($V_{\rm r}$) of the binary. Although the \gaia archive provides radial velocity estimates for some sources, these cannot be used for our study. For close binary systems, such as those in our sample, the systemic radial velocity we require is that of the centre of mass. By contrast, \gaia currently provides only an estimate of the radial velocity of the optically luminous component, which will be affected --- and usually dominated --- by its orbital motion. We have therefore instead compiled $V_\mathrm{r}$ estimates from the literature. These are available for 28 of our systems (10 \sg, 12 \be, 1 RLO, and 5 of unclear classes) as shown in Table~\ref{astrometric DR2}. When multiple radial velocity measurements were available in the literature, the most recent value based on detailed analysis was preferred, as this usually represented improvements resulting from use of modern instrumentation and revisitation of older observations leading to reduced uncertainties. For the remaining sources without literature $V_{\rm r}$ values, we first estimate the 2-D \vpec\ ($V_{\rm pec,2D}$) in the plane of the sky by assuming that the radial velocity is entirely associated with Galactic rotation. We then assume that the peculiar velocities are distributed isotropically, which allows us to estimate the 3-D \vpec\ ($V_{\rm pec,3D}$) by applying a correction factor of $4/\pi$. This factor is the ratio of the expectation values of the $V_{\rm pec,3D}$ and $V_{\rm pec,2D}$ values ($\langle V_{\rm pec,3D}\rangle/\langle V_{\rm pec,2D}\rangle$) for an isotropically distributed sample \citep{hobbs_statistical_2005}. Additional tests of the isotropic assumption will be described in Section~\ref{Isotropy}.

 In our calculations, we first calculate velocities in an equatorial Cartesian system, following \citet{perryman_hipparcos_1997}. We then convert these components to a Galactic Cartesian system. Solar motion and circular Galactic rotation are removed following \citet{reid_trigonometric_2009}, yielding the desired \vpec, which are relative to our sources' expected motion in the Galactic plane \citep{gandhi_orbital_2020}. The Cartesian components of solar motion relative to the local standard of rest ($U_{\sun}$, $V_{\sun}$, and $W_{\sun}$ are 8.0 $\pm$ 0.9, 12.4 $\pm$ 0.7, and 7.7 $\pm$ 0.9 \kms, respectively) and Galactic rotation speed at Solar distance ($\Theta_{0} = 236 \pm 3$ \kms) from the Galactic centre ($R_0 = 8.2 \pm 0.1$ kpc) are taken from the work of \citet{kawata_galactic_2019}. We estimate uncertainties via Monte Carlo resampling of all relevant parameters, including the Galactic constants. When quoting uncertainties, we adopt the highest-density interval (HDI)\footnote{\url{https://github.com/aloctavodia/BAP/blob/master/first_edition/code/Chp1/hpd.py}} containing 68.27 per cent of the Monte Carlo samples. For systems without available $V_{\rm r}$ measurements with the isotropic correction of $4/\pi$, an additional uncertainty term of 0.18 dex in log\vpec\ was included. This term was determined from simulations of isotropically distributed velocities and accounts for the scatter introduced when converting from 2D to 3D velocities.
 (e.g., \citealp{blaauw_origin_1961}) 
 
 It is important to emphasize that the true space velocities of HMXBs can only be accurately determined if the system’s birthplace is known (for example, the origin of 4U 1700--37 in NGC 6231; see \citealp{van_der_Meij_2021}). Without knowledge of the birthplace, velocities are best estimated relative to the Local Standard of Rest (LSR). However, massive stars in the Galactic plane commonly exhibit deviations of about 20 \kms from the LSR \citep{carlberg_1985}, implying that the velocities we derive may not precisely represent their true three-dimensional motions.

\section{Results}\label{sec:result}

\begin{figure}
	\includegraphics[width=\columnwidth]{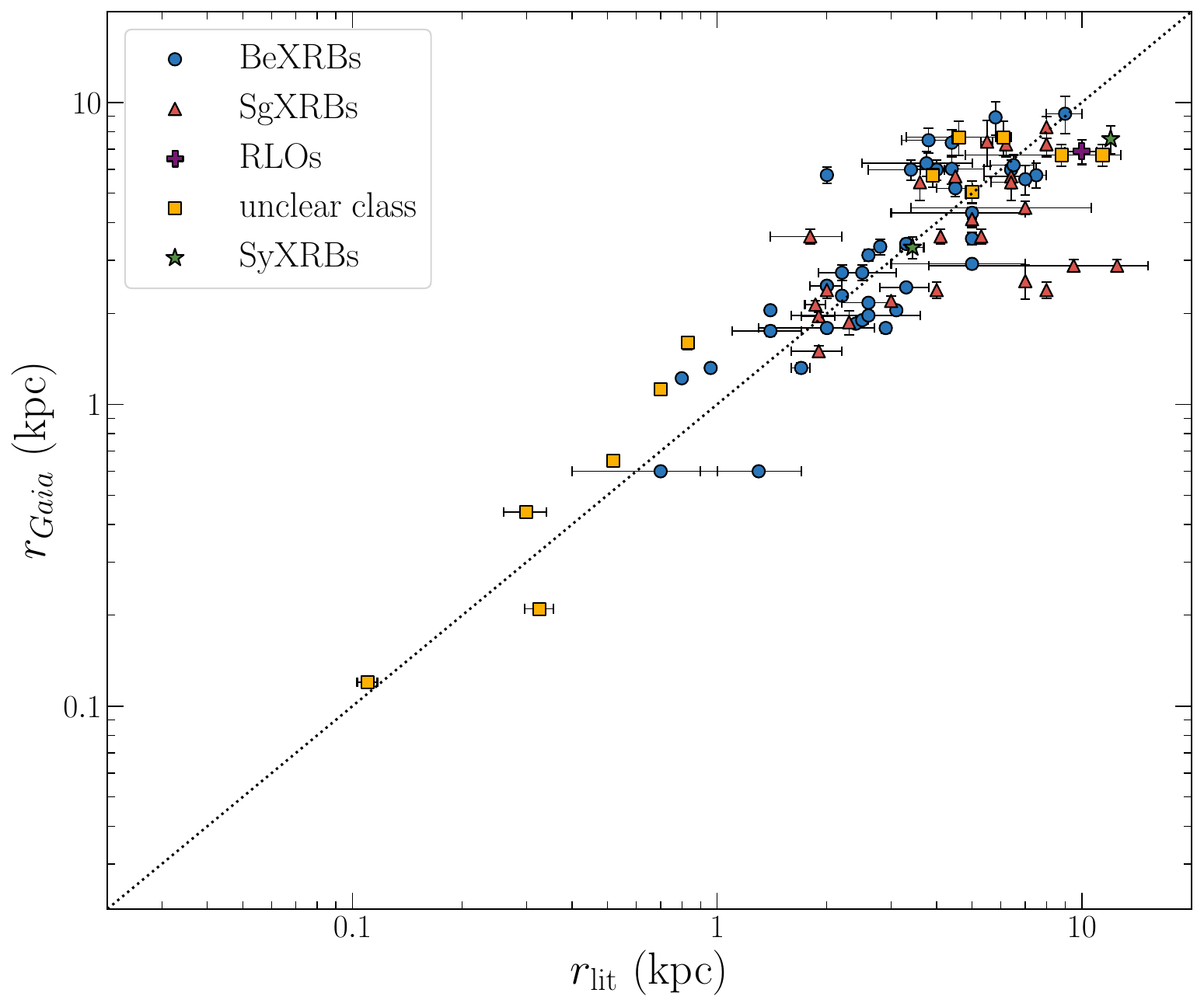}
    \caption{Comparison of \gaia DR3 distances ($r_{\gaia}$) with literature estimates ($r_{\rm lit}$) for our HMXBs, where known. The dotted line denotes the equality relation $r_\mathrm{lit} = r_\gaia$. Symbols represent different spectral classes. Sources scatter around the equality relation, and there is no obvious bias as a function of spectral class.} 
    \label{distance}
\end{figure}

Fig.\,\ref{distance} compares our \gaia-based distances ($r_{\gaia}$) with literature values ($r_{\rm lit}$), overlaid with a dotted line denoting the 1:1 relation ($r_{\gaia} = r_{\rm lit}$) and using symbols to indicate different spectral classes. The agreement between the parallax-based and literature estimates is generally reasonable, despite some scatter, and there are no obvious systematic dependencies on spectral class. In principle, the \gaia geometric parallax estimates should be more reliable than literature distances, which often rely on heterogeneous, model-dependent methods. Some HMXBs in our sample have more than one reported distance, although not all of these have associated uncertainties. Our distance estimates and their uncertainties are listed in Table~\ref{d}, with notes on individual sources provided later, and we find that the \gaia distances derived via parallax inversion agree with the Bayesian estimates reported for HMXBs by \citet{Zhao_2023}. To further investigate the residual scatter in the figure, we checked the values of their Gaia-reported astrometric excess noise and Re-normalised Unit Weight Error (RUWE). These parameters are indicative of residuals relative to the Gaia pipeline single-star astrometric fits. Values significantly in excess of 0 for the astrometric excess noise, and values in excess of 1.4 for RUWE could indicate a poor single-star fit, which may result from instrumental or pipeline artefacts, or arise from the presence of inherent stellar multiplicity (cf. \citealt{belokurov20, gandhi_aen_2022}). Our targets are specifically selected to be binaries and, indeed, we did not find any systematic trends with RUWE or astrometric excess noise. This implies that even if single-star fitting accounts for some the scatter between $r_\gaia$ and $r_\mathrm{lit}$, there is no obvious bias for individual source distances.

\begin{figure}
    \includegraphics[width=\columnwidth]{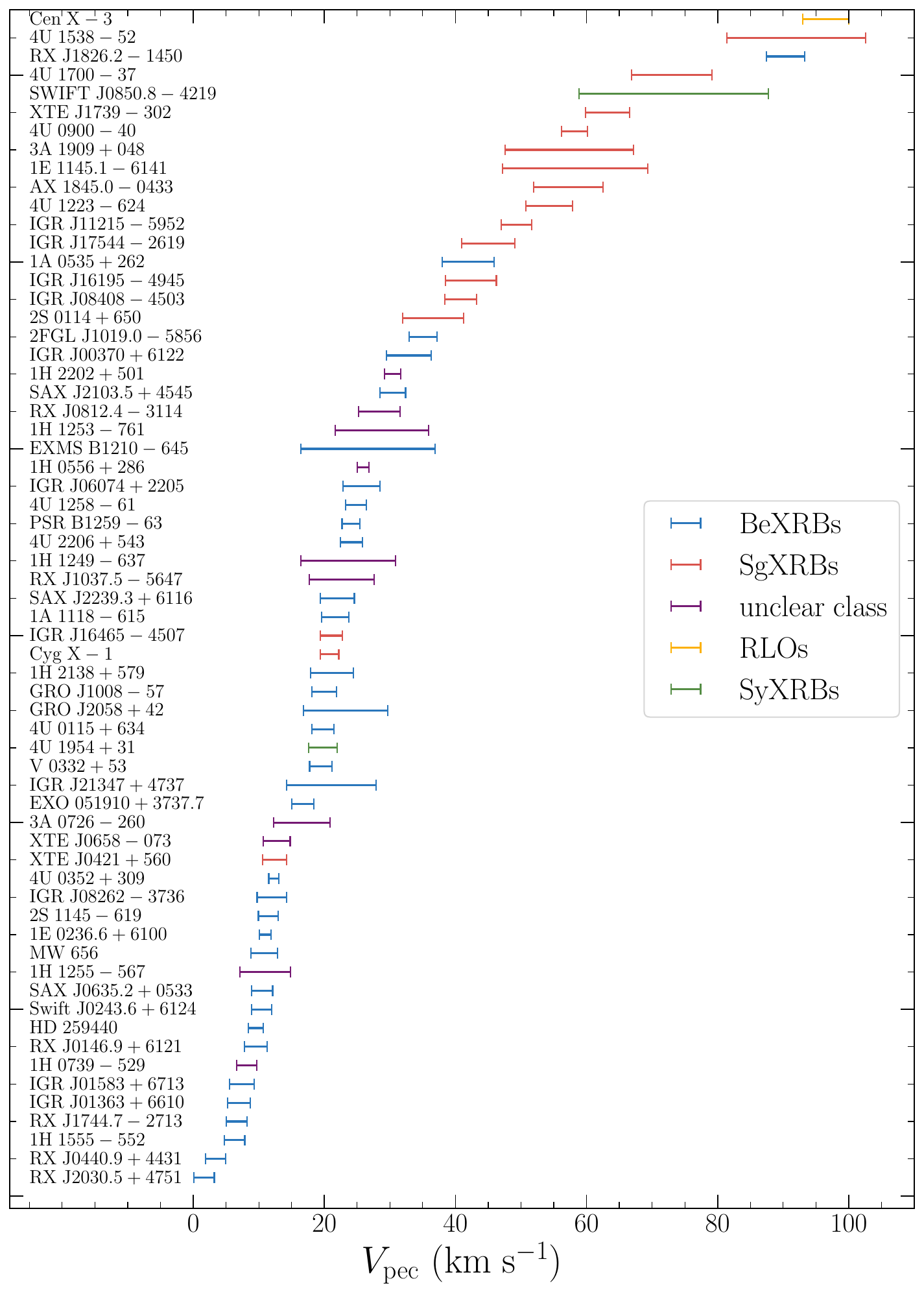}
    \caption{Highest-density (68.27 per cent) intervals of \vpec\ for 63 HMXBs. Source names are annotated on the left. Sources are sorted by \vpec\ in ascending order from the bottom, and are colour-coded by HMXB class.}
    \label{vp_ordered}
\end{figure}

Source kinematics were computed using the algorithm described in Section~\ref{sec:vpeccomputation}. In addition, by generating 50,000 mock samples through Monte Carlo resampling of all parameters, including the adopted Galactic constants, we derived  probability distributions of \vpec\ for all systems. These distributions, categorised by their respective sub-groups, are presented in Appendix Fig.\,\ref{vp_2class_dis_inc_unk_identity} and \ref{vp_2class_dis_inc_unk_identity-rlo_sy_unc}.

Fig.\,\ref{vp_ordered} is a summary of these \vpec\ confidence regions for each source. Sources are sorted by their \vpec\ values and colour-coded by HMXB sub-classes. It is easily apparent from the figure that SgXRBs have larger \vpec\ values than BeXRBs. The estimated \vpec\ values of HMXBs are tabulated in Table~\ref{d}, with a full range spanning $\approx$\,2–-97 \kms and an average of 29.0 $\pm$ 2.8 \kms, where the uncertainty represents the standard error. The single highest $V_{\rm pec}$ value is associated with the RLO system Cen X--3 ($V_{\rm pec}$ = $96.5\pm3.5$ \kms). However, this is the only RLO HMXB in our sample. More data will be needed to determine if there is a tendency for RLO systems to be fast movers. 

Focusing on the two more sizeable sub-samples, we find that the mean \vpec\ values for \be and \sg are 20.9$\pm$3.5 and 58.0$\pm$6.6 \kms, respectively. This difference persists even with the restricted subset of HMXBs that have measured $V_{\rm r}$ values. For reference, the velocity dispersion of young stellar populations is approximately 20 \kms \citep{carlberg_1985}.

\begin{figure}
    \includegraphics[width=\columnwidth]{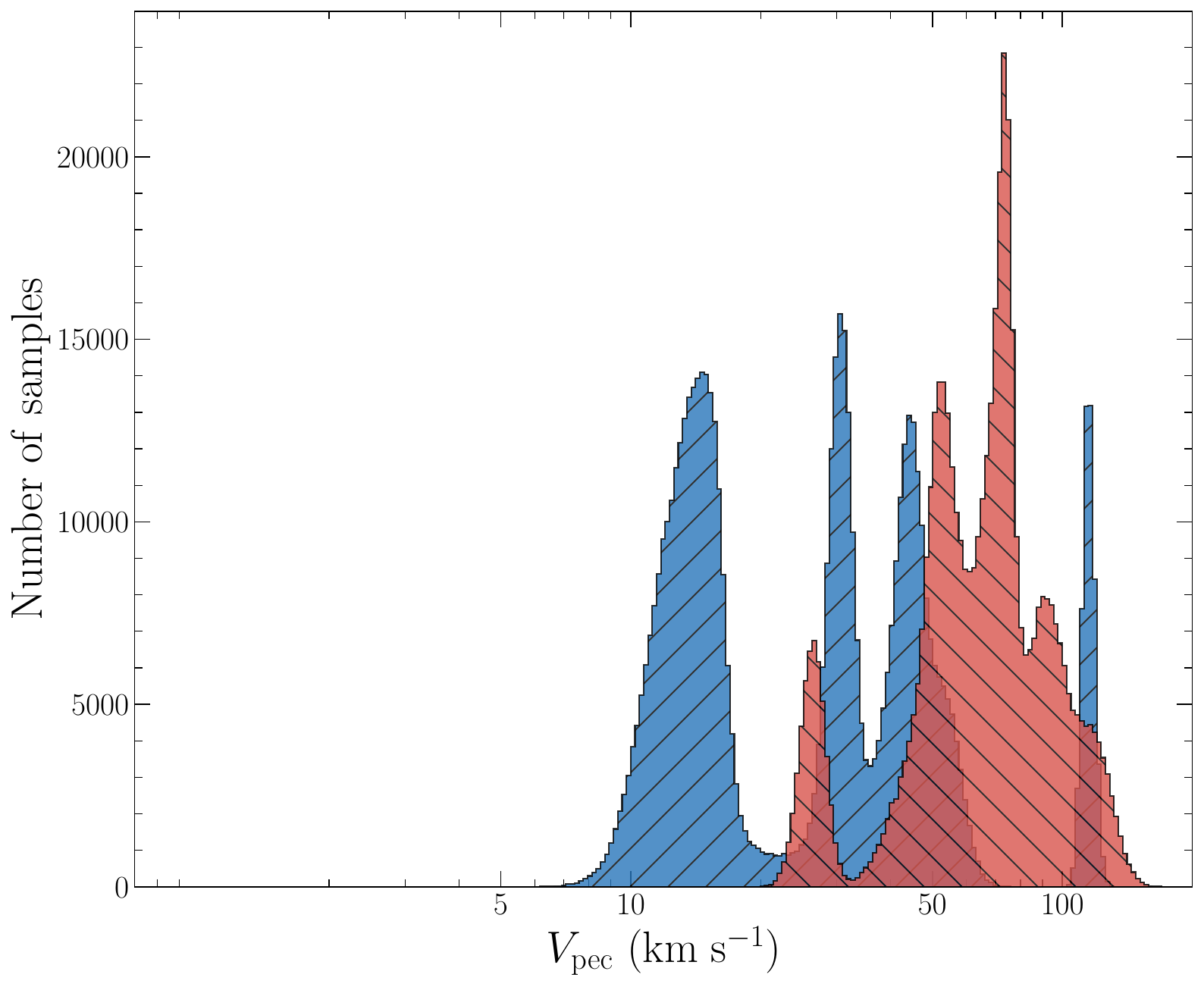}
    \caption{Summed distributions of $V_{\rm pec}$ for \be\,(blue, right-angled hatching `/') and \sg\,(red, left-handed hatching `$\setminus$'; colours online). Only systems with available $V_\mathrm{r}$ values are included here. For each source, 50,000 random samples are drawn. The \be\ system with the highest \vpec\ velocity is RX J1826.2--1450/LS 5039 -- its distribution stands out from its subgroup on the far right. Similarly, Cyg X--1 is the SgXRB with the smaller \vpec, $\approx$ 20 \kms, separated from other SgXRBs.
    }
    \label{vp_2class_dis}
\end{figure}

We illustrate this difference between these two classes again with the histogram in Fig.\,\ref{vp_2class_dis}. Here, 50,000 values for each of these HMXBs were randomly sampled from their respective \vpec\ distributions. These sample distributions are then summed for the BeXRBs and SgXRBs separately. Amongst the BeXRBs, RX\,J1826.2--1450/LS 5039's distribution appears to be an outlier. Similarly, Cyg X-1 has a low $V_{\rm pec}$ ($\approx 20$ \kms) compared to other SgXRBs. In the Appendix, the reader can find larger figures with individual systems annotated.

\begin{figure}
    \includegraphics[width=\columnwidth]{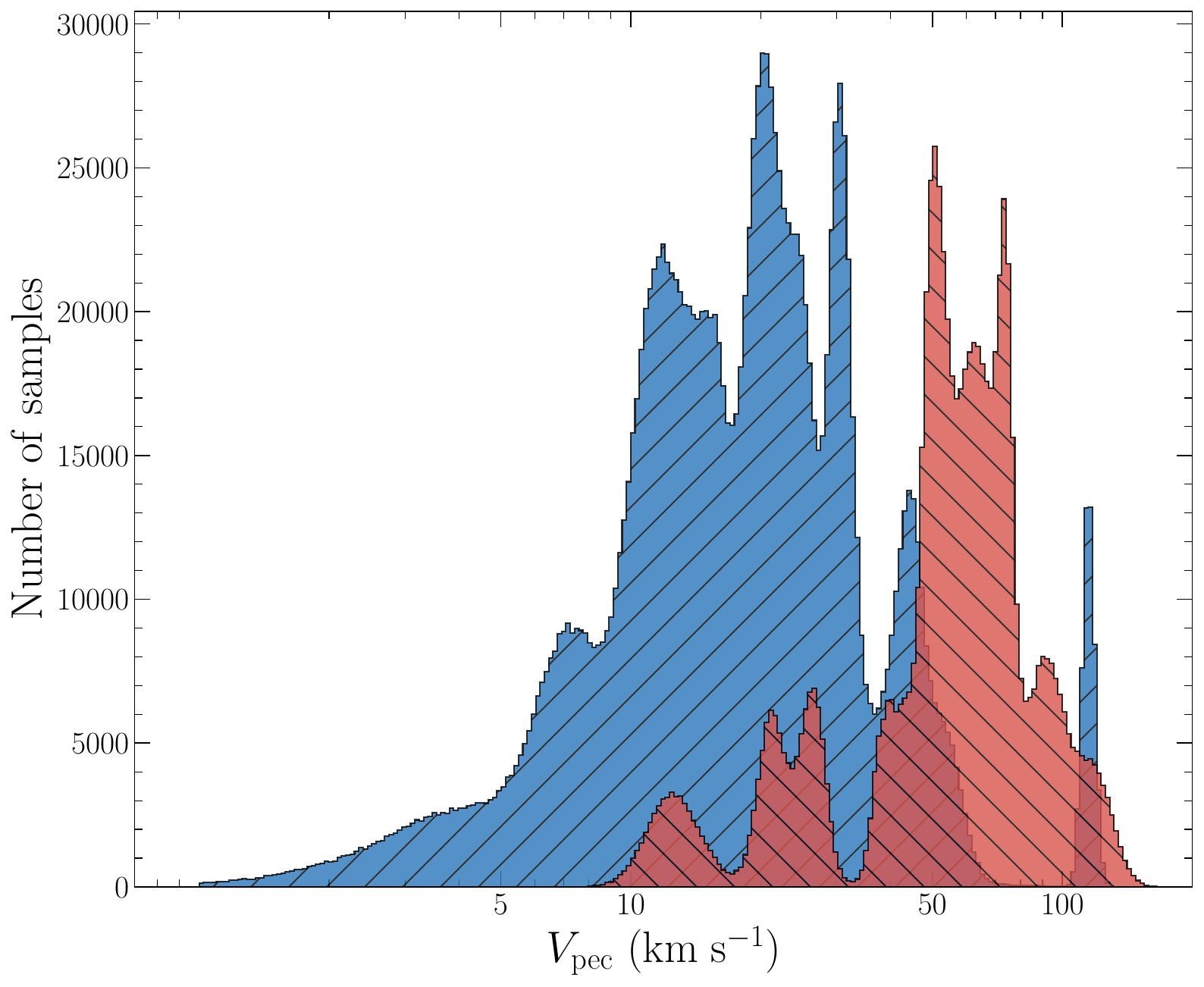}
    \caption{Histogram of \vpec\ values as in Fig.\,\ref{vp_2class_dis}, but now also incorporating $V_{\rm pec,3D}^{\rm iso}$ as a proxy for systems without established radial velocities.}
    \label{vp_2class_dis_inc_unk}
\end{figure}

In Fig.\,\ref{vp_2class_dis_inc_unk}, we show the $V_{\rm pec}$ distribution for all 49 systems with or without measured $V_{\rm r}$ values.
Both Figs.\,\ref{vp_2class_dis} and \ref{vp_2class_dis_inc_unk} clearly demonstrate a significant difference in the mean velocities between \be\ and \sg. Including or excluding systems without measured radial velocities does not change this inference, pointing to a robust difference between the classes.

To quantitatively assess this difference, we performed a Kolmogorov-Smirnov test (K-S test) using the {\tt ks\_2sample} package within the {\sc scipy}\footnote{\url{https://scipy.org/}} library \citep{scipy}. The null hypothesis is that the \vpec\, values for \be\, and \sg\, are drawn from the same underlying distribution. 
We created 1,000 random ensembles of \vpec\ by drawing one random sample from the \vpec\, distributions of each system.
The K-S test is then performed on each of the 1,000 random ensembles, from which we obtained a distribution of test statistics and $p$-values. We found that 100 per cent of the test results reject the null hypothesis, implying a significant difference between the \sg\, and \be\, \vpec\ distributions. 

\begin{figure}
    \includegraphics[width=0.987\columnwidth]{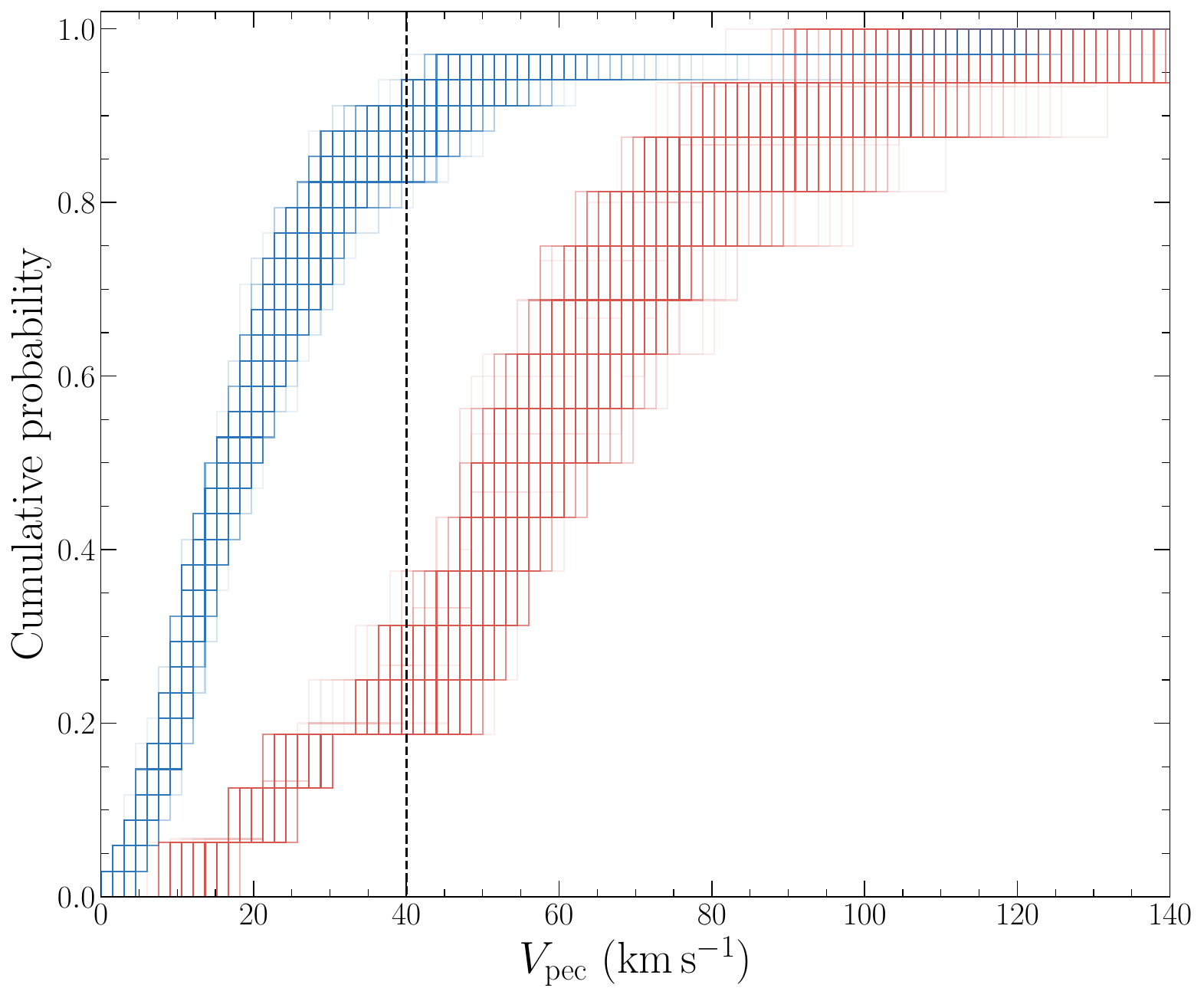}
    \caption{Cumulative distribution of $V_{\rm pec}$ for \be\,(blue) and \sg\,(red). Roughly half of the \be\, have $V_{\rm pec}$ lower than 20 \kms, with nearly all of them being slower than 30 \kms. In contrast, around half of \sg\, have $V_{\rm pec}\leq 50$ \kms. The two subgroups are optimally separated around a threshold of 40 \kms, represented by the vertical dashed line.}
    \label{cdf_hist}
\end{figure}

In Fig.\,\ref{cdf_hist}, the cumulative distribution shows that around 50 per cent of \be\, have \vpec\, below 20 \kms, and 80 per cent of them are slower than 30 \kms. On the other hand, half of the \sg\, sample exhibits \vpec\, values exceeding 50 \kms. The two sub-groups are maximally separated at a velocity of approximately 40 \kms. Accordingly, we adopt this value as our velocity separation threshold: systems with \vpec\, equal to or greater than 40 \kms are classified as having \textit{high} peculiar velocities, while those with \vpec\, less than 40 \kms are classed as \textit{low} peculiar velocity systems. Although \textit{low} can be relative, a space velocity of less than 40 \kms\ is still substantially higher than the typical sound speed in most regions of the Galactic plane. Interestingly, three-quarters of \sg\, exhibit \vpec\, exceeding this threshold.

\begin{figure*}
    \includegraphics[width=0.65\textwidth]{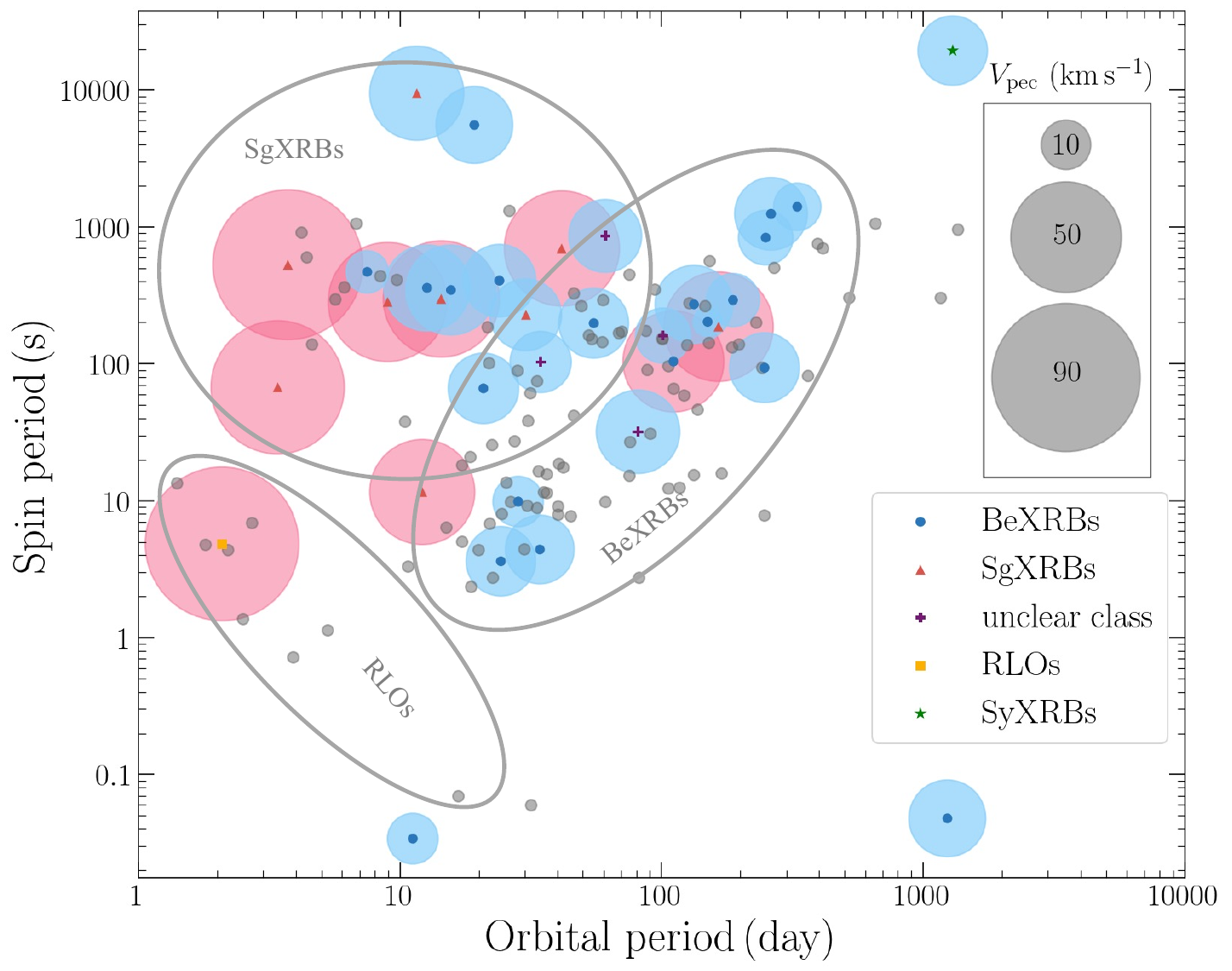}
    \caption{The Corbet diagram (colours online) plotting spin vs. orbital period for HMXBs in this study; different systems are further distinguished by the colour of the crosses: blue, BeXRBs; red, SgXRBs; Magenta, unclear classes; yellow, RLOs. The \vpec\ values are mapped to the sizes of circles around the crosses. Red and blue circles indicate systems with \vpec\ greater and lower than 40 \kms, respectively. Additionally, filled grey circles mark the loci of other Galactic and extragalactic HMXBs from Small Magellanic Cloud, and Large Magellanic Cloud.}
    \label{corbet}
\end{figure*}

There are 33 systems that have available orbital and spin periods, listed in Table \ref{d}. To investigate any potential associations between \vpec\ and the spin or orbital periods, we visualise the distribution of these periods in Fig.\,\ref{corbet}. This shows a Corbet diagram \citep{corbet_three_1986} for these systems, with \vpec\ values depicted by symbol size. It is immediately apparent that sources also show some degree of separation in the Corbet parameter space. These \vpec\, values and corresponding implications will be discussed in Section~\ref{sec:discussion}.

\section{Discussion}\label{sec:discussion}

The birth of compact objects is expected to leave an imprint on their subsequent kinematics and evolution. These can be constrained in binaries where the companion star traces the orbit of the system. In this work, we have leveraged state-of-the-art astrometry from {\em Gaia} DR3 to measure the complete (three-dimensional) systemic motions of Galactic HMXBs in excess of Galactic rotation that have been perturbed by natal kicks. Our results presented in Section~\ref{sec:result} quantify the moments of the three-dimensional kinematic distributions for the full sample, and reveal differences between HMXB classes; we extend previous studies that were either restricted to two-dimensional velocities or could not clearly reveal such differences \citep{chevalier_hipparcos_1998, fortin_constraints_2022}. Recent studies of larger samples including black holes as well as neutron stars in binaries find wider \vpec\ distributions extending to several hundred \kms\ (e.g. \citealt{Zhao_2023}); our work here has focused on the more massive of such systems, which tend to have lower peculiar motions, on average, presumably also indicative of weaker corresponding natal kicks. But we also hone in on the various sub-classes of HMXBs to explore their properties in more detail than before. 

\begin{figure}
    \includegraphics[width=\columnwidth]{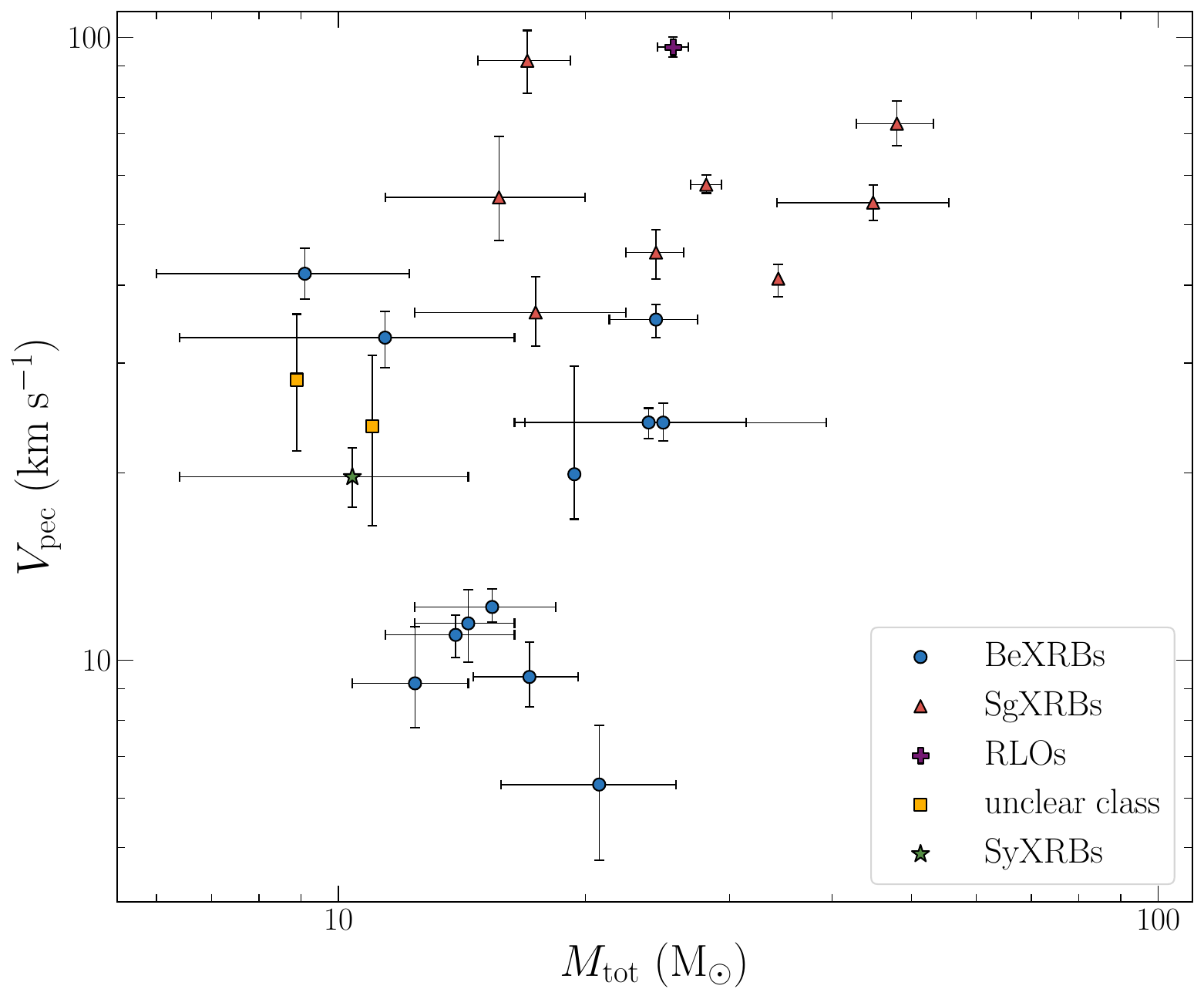}
    \caption{The relationship between total mass ($M_{\rm tot}$) and \vpec\ for NS HMXBs, including their associated uncertainties. Systems are distinguished by cross colour according to sub-classes: blue for BeXRBs, red for SgXRBs, green for SyXRBs, magenta for unclear classifications, and yellow for RLOs.}
    \label{mtot}
\end{figure}

A global anti-correlation between total mass and \vpec\ across all types of compact object binaries -- including HMXBs {\em and} LMXBs -- has previously been identified by \citet{Zhao_2023}. Since $M_{\rm tot}$ is the primary difference between LMXBs and HMXBs, this trend {\em across} classes need not necessarily apply {\em within} classes. In Fig.~\ref{mtot}, we therefore show what this parameter space looks like when we focus specifically on HMXBs. Formally, a Spearman rank test yields a marginally significant {\em positive} correlation for these systems (correlation coefficient = 0.46, $p$-value = 0.02). To the extent that the correlation is real, Fig.~\ref{mtot} suggests that it is induced by systematic differences in mass and \vpec\ between different HMXB sub-types. Given the unclear statistical significance of this trend, we will not discuss it further in the present study. However, it would clearly be worth revisiting this topic when additional data become available.

We now discuss some of the implications of our results. 

\subsection{Comparison of kinematics between \sg\ and \be} \label{sec:compare_be_sg}

What is the underlying cause of the difference between the subtypes? One possibility can be traced back to differences in the pre-SN progenitors of their compact objects. SgXRBs are thought to inhabit binary systems with tighter orbits, on average, as compared to BeXRBs. They are thus expected to have higher relative orbital velocities between the two binary components at the point of SN. Following the kinematic formulation of \citet{kalogera_orbital_1996}, we expect the average runaway velocity to be of a similar order to the relative orbital velocity, and thus be larger for SgXRBs. 
 
This scenario was first pointed out by \citet{van_den_heuvel_origin_2000}. Their study proposed that the kinematic differences between Be and Sg system may be tracable to two main factors. The first factor is related to a higher probable fractional helium core mass of the progenitor stars of \sg\, compared to those of \be. A higher primary mass for \sg\ leads to a higher helium core mass and, in turn, a smaller increase in the orbital period during the initial mass transfer from the primary to the companion, resulting in tighter pre-supernova (pre-SN) orbits and higher orbital velocities for the helium core. The second proposed factor is a proportionally smaller mass ejection in Be systems compared to supergiants during the SN event. We will return to these points in Section \ref{sec:binpop}. 

\citet{fortin_constraints_2022} confirm that \be\ possess systems with relatively low mass and low peculiar velocities. Furthermore, in one BeXRB system at the extreme end of the \be\ mass scale, RX J1826.2--1450/LS 5039, hosting an Oe-type donor, the runaway velocity $V_{\rm pec}$ is also extreme, at $\approx$ 89.1 \kms. By contrast, the other Oe-type objects studied display $V_{\rm pec}$ values between 20 and 40 \kms. On the other hand, the \sg\, display a consistently high peculiar velocity, with little to no relationship to mass \citep{fortin_constraints_2022}. 

For systems in which the accretor is a neutron star, it is interesting to check whether/how the peculiar velocity of systems relates to their position in the so-called Corbet diagram, i.e. to their orbital and spin periods ($P_{\rm orb}$ and $P_{\rm spin}$, respectively). Fig.\,\ref{corbet} shows this diagram with both sub-classes and peculiar velocities added. This confirms that \vpec\ can provide useful supplementary information for classifying HMXBs. Nonetheless, it should be noted that there are exceptions. That is, position in the Corbet diagram does not {\em cleanly} correlate with either HMBX sub-class, nor with \vpec. Nevertheless, the difference in the characteristic peculiar velocities of BeXRBs and SgXRBs, in particular, is also clearly reflected in the Corbet diagram. 

\citet{Knigge_2011} showed that BeXRBs appear to fall into two distinct sub-populations: systems with short $P_{\rm spin}$ ($\lesssim 30$~s) and systems with long $P_{\rm spin}$ ($\gtrsim 30$~s). They suggested that the low-$P_{\rm spin}$ group may originate from low-kick electron-capture supernovae, in which case they should exhibit lower space velocities than those with long $P_{\rm spin}$. We have attempted to test this suggestion. Within our set of BeXRBs, a K-S test reveals no statistically significant difference between these two $P_{\rm spin}$ groups as indicated by the very small $p$-value. However, there are only five systems in the short spin-period category in our sample, so this test has very limited statistical power. We can therefore not conclusively confirm or reject this possibility.

\subsection{Isotropy of Peculiar Motions} \label{Isotropy}
Analysis of the full 3-D peculiar motions of sources  requires knowledge of $V_{\rm r}$. However, $V_{\rm r}$ values are known only for 28 of our sample systems. For the remaining systems, we have made the assumption of isotropicity. This is the minimal ansatz that applies if there is no special directionality to the motions of HMXBs in the Galactic potential, as seen by us. Here, we conduct a basic test on the viability of this ansatz. 

Such an ansatz was first introduced for compact object motions in the pioneering study of \citet{hobbs_statistical_2005} who utilised the 2-D proper motions of pulsars to infer the properties of 3-D velocity distributions. 
In a statistical sense, the expected $\langle V_{\rm pec,3D}\rangle/\langle V_{\rm pec,2D}\rangle$ for a Maxwell distribution of velocities is expected to be 4/$\pi$. We can test this directly for sources with full 3-D kinematic information. 

\begin{figure}
    \includegraphics[width=\columnwidth]{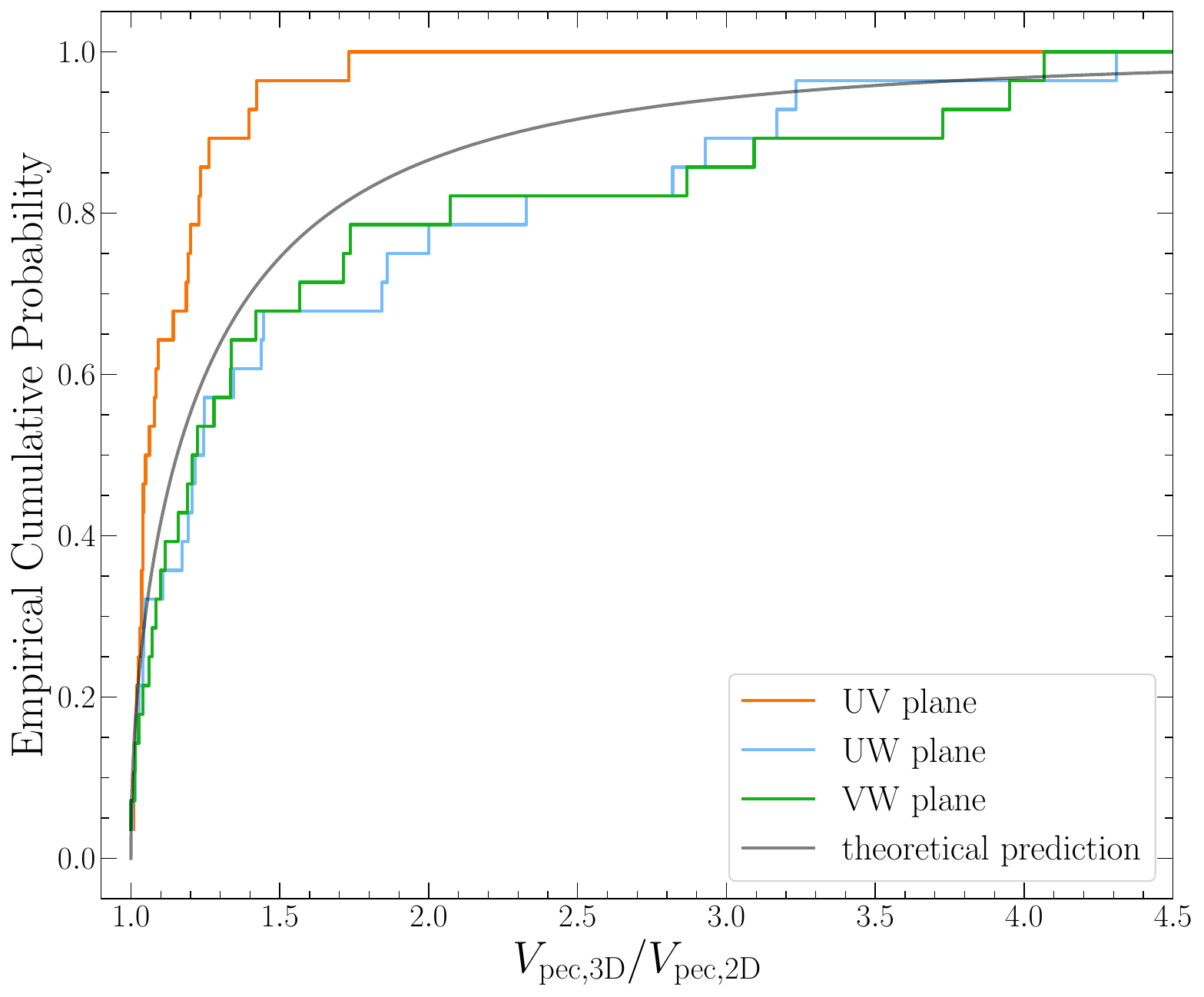}
    \caption{The empirical cumulative distribution of the ratio between 3-D and 2-D of \vpec. There are three possible cases for $V_{\rm pec,2D}$ ($V_{\rm pec,2D}$ on UW, UV, and VW plane) and 10 million theoretical predictions. The distribution suggests that when using $V_{\rm pec,2D}$ from the UW and VW planes, the resulting ratios align closely with the theoretical predictions. Furthermore, results from the K-S test, $p$ value, indicate that $V_{\rm pec,2D}$ on the UV plane shows no significant deviation from the theoretical predictions.}

    \label{ecdf_UVW}
\end{figure}

For the 28 objects with measured $V_{\rm r}$, we can compute $V_{\rm pec,2D}$ by treating them as if their radial velocities are unknown. The 3-dimensional $V_{\rm pec,3D}$ represents the resultant vector of space velocity in Cartesian coordinates ({$U_{\rm s}$}\footnote{$U_{\rm s}$: Radial component of $V_\mathrm{pec}$ in the Galactic plane}, {$V_{\rm s}$}\footnote{$V_{\rm s}$: Azimuthal component of $V_\mathrm{pec}$ relative to local Galactic rotation}, and {$W_{\rm s}$}\footnote{$W_{\rm s}$: $V_\mathrm{pec}$ component perpendicular to the Galactic plane}), resulting in three possible cases for $V_{\rm pec,2D}$ (UW, UV, and VW plane). The calculated $\langle V_{\rm pec,3D}\rangle/\langle V_{\rm pec,2D}\rangle$ values in each plane are shown in Table \ref{isotropy_table}. For these three cases, closely aligning with the expected ratio of 4/$\pi$, validating the underlying assumption. We compared the ratio between $V_{\rm pec,3D}$ and $V_{\rm pec,2D}$ with the theoretical prediction from a simulation which is drawn from 10 million random samples as shown in Fig.\,\ref{ecdf_UVW}. We also tested the differences between the theoretical prediction and our estimated values using the K-S test. For the UW and VW planes, the test results indicate strong consistency between empirical data and simulations. However, for the UV plane, the K–S test yielded a $p$-value of 0.06—marginally above the conventional significance threshold of 0.05—suggesting only tentative agreement and highlighting the need for further investigation into this particular component.

\begin{figure*}
    \includegraphics[width=0.8\textwidth]{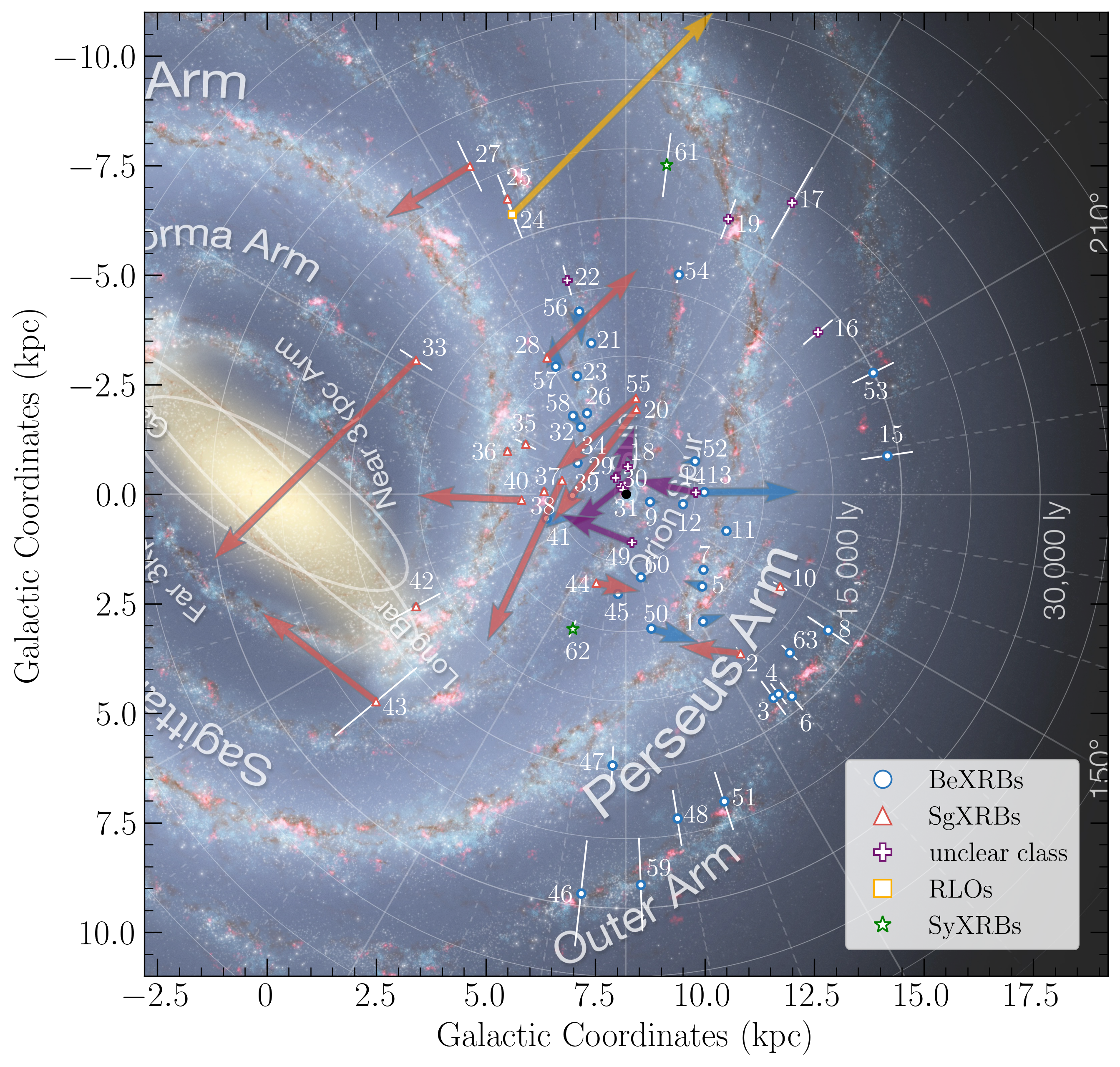}
    \caption{A map of the Galactic plane with the projected locations of 63 HMXBs, incorporating their distance uncertainties accounted for through error propagation (Milky Way image courtesy of NASA/JPL-Caltech, ESO, J. Hurt). The blue dots represent BeXRBs, red dots represent SgXRBs, yellow dot represents RLOs,  magenta dots represent unclear classes, green dots represent SyXRBs, and the black dot at the centre is the Sun. The length and direction of the arrows indicate $V_\mathrm{pec,2D}$ on UV plane ($V_{\rm pec}$ projected on the Galactic plane) for 28 systems with available $V_{\rm r}$ estimates. Numeric labels correspond to system numbers listed in the accompanying table.}
    \label{lon_distance}
\end{figure*}

Fig.\,\ref{lon_distance} shows the Gaia DR3 map displaying the positions of 63 HMXBs projected onto the Galactic plane, taking into account distance uncertainties and the different colours representing 4 sub-classes and unclear class. The 28 arrows displayed on the plot represent the $V_{\rm pec,2D}$ vectors on the UV plane, indicating both their length and direction for 28 known $V_{\rm r}$ sources. Notably, the distribution of the $V_{\rm pec,2D}$ vectors on the UV plane suggests isotropy, indicating a uniform and consistent pattern across the Galactic plane.

\begin{table}
    \centering
    \caption{The calculated $\langle V_{\rm pec,3D}\rangle/\langle V_{\rm pec,2D}\rangle$ values for each plane and the $p$-values from the K-S test compared to the theoretical prediction under the assumption of isotropy.}
    \label{isotropy_table}
    \begin{threeparttable}
    
    \begin{tabular}{ccc} 
		\hline

		$V_{\rm pec,2D}^a$	& 			$\langle V_{\rm pec,3D}\rangle/\langle V_{\rm pec,2D}\rangle^b$ 	&  $p$-value$^c$  \\
		\hline
UV plane  &   1.20    &   0.06  \\
UW plane &   1.33    &   0.19   \\
VW plane &   1.40    &   0.45   \\
Theoretical prediction &   4/$\pi$ (1.27)   &  -- \\
assuming isotropy & & \\  
          \hline
    \end{tabular}  
    \begin{tablenotes}
    \footnotesize
    \item[] \footnotesize{$^a$: The considered plane of $V_{\rm pec,2D}$} \\
    \item[] \footnotesize{$^b$: The ratio between the average $V_{\rm pec,3D}$ and the average $V_{\rm pec,2D}$} \\
    \item[] \footnotesize{$^c$: $p$-value resulting from the comparison between each plane and 10 million theoretical predictions were obtained using K-S test} \\
    \end{tablenotes}

    \end{threeparttable}
\end{table} 

Based on the discussion mentioned above, we assume that the $V_{\rm pec,2D}$ would be the smallest magnitude relative to the Galactic rotation. Therefore, $V_{\rm pec,min}$ can be considered as 2-D speeds. In order to estimate 3-D speeds, we use the assumption mentioned above (i.e., isotropy of the velocity vector) multiplied by a constant, referred to as $V_{\rm pec,3D}^{\rm iso}$. The $V_{\rm pec,3D}^{\rm iso}$ values in this study correspond with \citet{fortin_constraints_2022}. Therefore, we use this estimation to assume $V_{\rm pec,3D}^{\rm iso}$ for systems with no established $V_{\rm r}$. Even though, there are ambiguously estimated $V_{\rm pec,3D}^{\rm iso}$ values for systems with no literature $V_{\rm r}$, we can still see the different velocities between \be\ and \sg\ as shown in Fig.\,\ref{vp_2class_dis}.

\subsection{Testing the Origin of Class Kinemetic Differences with Binary Population Synthesis}
\label{sec:binpop}

\citet{van_den_heuvel_origin_2000} suggest that the kinematic differences between the classes could arise from differences in progenitor properties -- specifically, the fractional mass-loss of the compact object progenitor through supernova (SN) and the pre-SN system orbital period. Testing this requires estimating the system properties at the time of SN. For this purpose, our team has begun to develop detailed simulations using  \textsc{`Compact Object Synthesis and Monte Carlo Investigation Code (cosmic)'}, a binary population synthesis code derived from the \textsc{Binary Stellar Evolution (bse)} framework, enhanced with updated evolutionary prescriptions and parameters \citep{Breivik2020}. Full details of our efforts will be presented in an upcoming work (Dashwood Brown et al., in preparation). One previous example case study on a black hole XRB -- H\,1705--250 -- outlines the most salient details of our methodology and can be found in \citet{dashwoodbrown24}.

In short, we simulate a large number of binaries, encompassing a broad range of initial parameters for the progenitor zero-age main sequence binary component masses, orbital period, and subsequent evolutionary pathways. In our simulations, stellar winds and mass transfer are treated according to \citet{Vink2001,Vink2005}; initial stellar metallicities range $0.1-2$\,Z$_\odot$; and we adopt a delayed SN mechanism, as outlined by \citet{Fryer2012}. Mass-loss from the SN progenitor star induces a kick to the centre-of-mass of the system \citep{Nelemans99}, and additional isotropic natal kicks are included for the resultant compact objects. These simulations are run to find binaries that survive the first SN and form an accreting compact object, searching for systems that end up matching the current observed properties of known XRBs. We match the current observed parameters (component masses, orbital period, and systemic \vpec) and, therefore, are able to estimate any key parameter of interest across the simulated ensemble. 

We note that some binaries are consistent with a broad range of pre-SN characteristics, and are sensitive to the natal kick prescriptions implemented. Conversely, in some instances, factors such as mass loss can be tightly constrained. By simulating a large ensemble of $>$\,10$^5$ systems, our methodology averages over the spread introduced by differing evolutionary pathways and unknown starting conditions. We can then use the distributions of physical parameters for successful simulations to extract the mean estimates of the likely pre-SN orbital periods and fractional mass-loss of progenitor stars for our XRBs. 

\begin{figure}
    \includegraphics[width=\columnwidth]{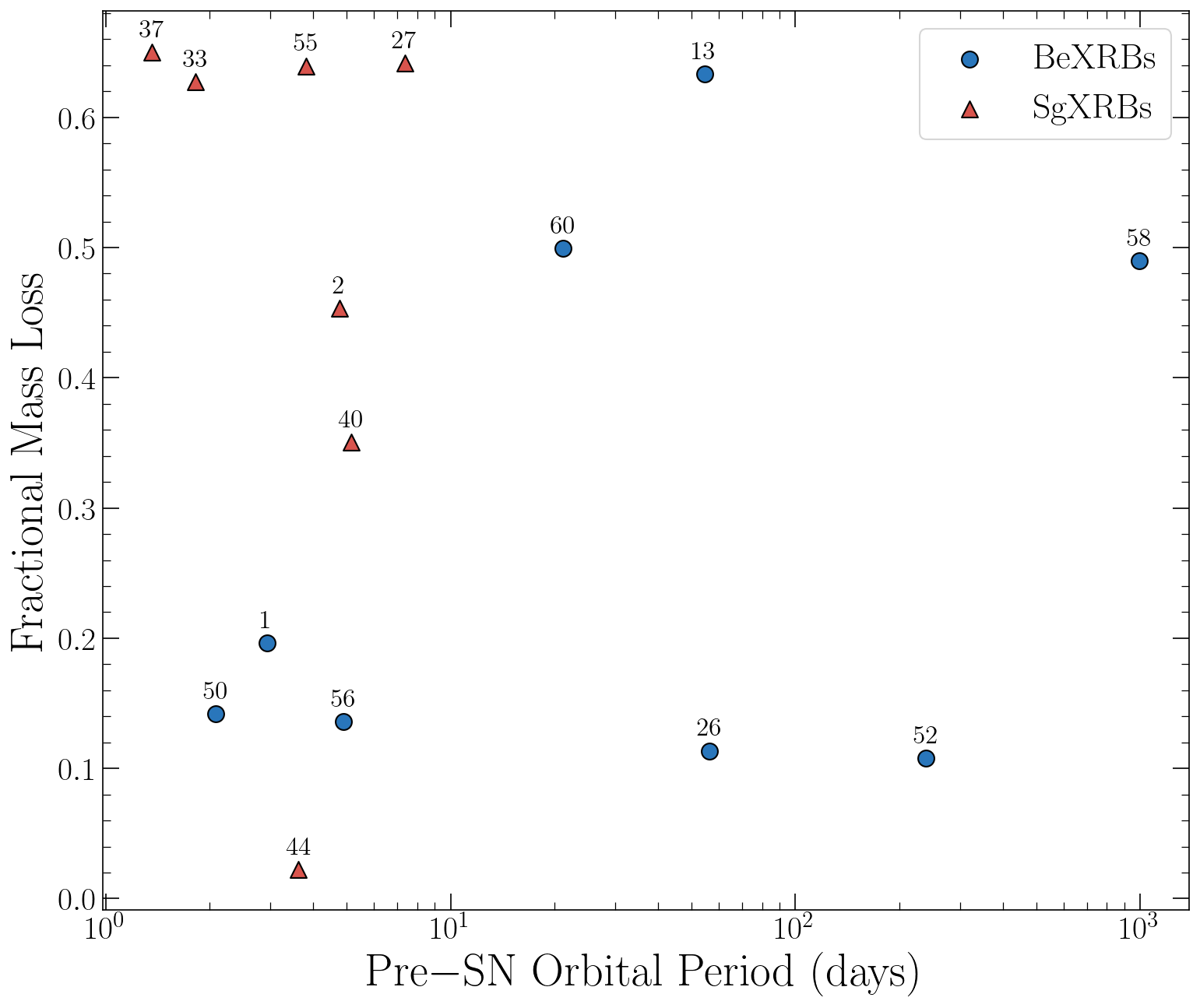}
    \caption{The relationship between pre-SN orbital period and fractional mass loss, based on our binary population synthesis simulations. Systems are distinguished by colour according to sub-classes: blue for BeXRBs, red for SgXRBs. Individual systems are annotated with their corresponding numbers from the Table.
    }
    \label{preSN-mass}
\end{figure}

Fig.~\ref{preSN-mass} illustrates the correlation between the mean pre-SN orbital period and fractional mass loss of the simulations for each of the plotted systems. There is a clear separation in how the classes are distributed between these parameters. The majority of \sg\ systems favour shorter pre-SN orbital periods, with a mean value of 4.0 days, and higher fractional mass loss, with a mean value of 0.5, compared to \be\ systems, which have mean values of 172.5 days and 0.3, respectively. These results, albeit preliminary, are aligned with the hypothesis propounded by \citet{van_den_heuvel_origin_2000} to explain the differences of the peculiar velocities between the classes. Further tests of this scenario should come from developing detailed stellar evolutionary calculations with {\sc mesa} \citep{mesa}. Updated astrometry from \gaia's new data releases (DR4, 5) in the future should also increase the sample of XRBs with robust kinematic measurements. 

\subsection{Completeness and Selection Effects}

According to \citet{neumann_xrbcats_2023}, there are currently 172 known HMXBs. Of these, 151 HMXBs have identified \gaia counterparts, but only 63 meet the stringent astrometric criteria set by \citet{bailer-jones_estimating_2015}. Consequently, the derived peculiar velocity (\vpec) values represent only 36 per cent of the confirmed HMXBs in our Galaxy, highlighting the potential impact of selection effects. In particular, the uncertainties in {\em Gaia} parallax measurements increase towards fainter magnitudes, ranging from 0.02–0.03 mas for \texttt{G} < 15 to 0.07 mas at \texttt{G} = 17 and 0.5 mas at \texttt{G} = 20 \citep{gaia_collaboration_gaia_2022}. Since our primary astrometric selection criterion is $\pi/\sigma_\pi > 5$, this effectively imposes an absolute-magnitude-dependent flux limit on our sample.  

\begin{figure*}
    \includegraphics[width=0.8\textwidth]{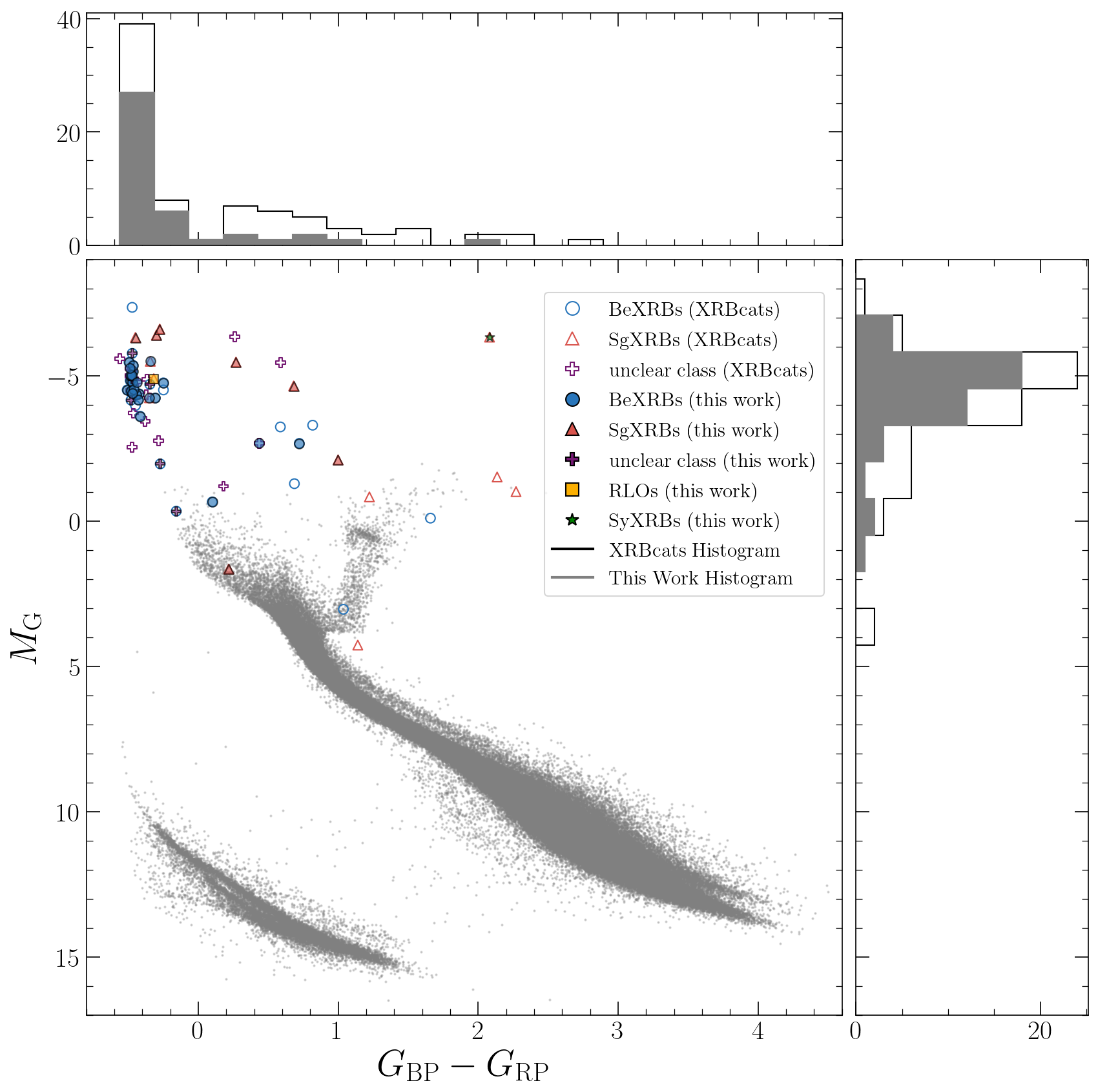}
    \caption{The colour-magnitude diagram (CMD) for HMXBs, where our targets are shown as filled symbols and XRBcats \citep{neumann_xrbcats_2023} as unfilled symbols. Sub-groups are represented as follows: circles for BeXRBs, triangles for SgXRBs, plus signs for unclear classifications, squares for RLOs, and stars for SyXRBs. Extinction and reddening corrections were applied, where available). The grey background dots represent 200,000 stars queried from the Gaia DR3 archive within 100 pc of the Sun \citep{gaia_collaboration_gaia_2018}. The top and right panels show histograms of absolute magnitude  $M_{\rm G}$  and colour index $G_{\text{BP}} - G_{\text{RP}}$, respectively, comparing our targets (filled histograms) with XRBcats (unfilled histograms).}
    \label{cmd}
\end{figure*}

In order to assess the potential impact of such selection effects on our sample, we present the colour-magnitude diagram (CMD) in Fig.~\ref{cmd}. Here, systems included in our sample are plotted with filled symbols, while the unfilled symbols show systems with \gaia counterparts in the XRBcats catalogue from \citet{neumann_xrbcats_2023}. Sub-groups are distinguished with different symbols: circles for BeXRBs, triangles for SgXRBs, plus signs for unclear classifications, squares for RLOs, and stars for SyXRBs. Extinction was accounted for in the plot, and systems without available extinction ($A_G$) and colour excess values ($E(Bp-Rp)$) were excluded. The grey background dots represent 200,000 stars from the Gaia DR3 archive within 100 pc of the Sun, selected following \citet{gaia_collaboration_gaia_2018}. Unsurprisingly, the optical counterparts of HMXBs typically occupy the bright and blue region in the CMD parameter space, given the early-type spectral classification of most systems. The CMD suggests that, if there is a systematic bias due to our sample selection, it seems to affect only the intrinsically faintest and reddest systems in XRBcats.

We can further examine the potential impact of selection effects by comparing the 1-D distributions in absolute magnitude and colour for our sample vs those for the XRBcats parent sample (see histograms at the top and right of Fig.~\ref{cmd}). No obvious differences are apparent between the two samples. We also construct an independent `control' sample of 32 sources (unfilled symbols in Fig.~\ref{cmd}) by retaining only those sources we are {\em missing} from XRBcats. We then {\em quantitively} compare our sample against this control sample via K-S tests on the absolute magnitude and colour distributions. These tests confirm that -- for both parameters -- the two samples are consistent with being drawn from the same underlying parent distributions. We therefore conclude that our sample should be fairly representative of the {\em known} HMXB population.

\section{Summary} \label{sec:summary}
We have investigated the peculiar velocity distribution of Galactic HMXBs by combining data from \gaia DR3 with literature $V_{\rm r}$ estimates. The salient highlights of our work can be summarised as follows:
    \begin{enumerate}
        \item A search was conducted within a 0.5\arcsec radius in \gaia DR3 to identify HMXB candidates, resulting in the detection of optical counterparts for a total of 63 systems with a maximum of parallaxuncertainty threshold of 20 per cent. These systems have \texttt{G}-band magnitudes ranging over 6 and 14, and are predominantly located close to the Galactic plane.
        \item The distribution of estimated distances based on \gaia parallaxes agrees with that based on literature distances, albeit with some scatter. No obvious trend or bias in distance estimates is found as a function of source class or astrometric fit quality. 

        \item The \vpec\, distribution is broad, with a mean velocity of $\approx$\,29 \kms\ and maximum values extending up to $\approx$\,100 \kms. The mean \vpec\, for BeXRB and SgXRB sub-groups are estimated to be 20.2 \kms\, and 48.9 \kms, respectively. Accounting for the scatter of stellar velocities in the background Galactic disc is expected to moderates these $V_{\rm pec}$ estimates, but will not impact the inference of a kinematic segregation between the source classes. 

        \item The overall \vpec\, distribution reveals two kinematically distinct sub-populations, centred around $\approx 40$ \kms. The low-velocity sub-population is predominantly associated with BeXRBs, while the high-velocity sub-population corresponds to SgXRBs.  

        \item The two SyXRBs within the HMXB population exhibit significantly different \vpec\, values from one another, indicating potentially distinct evolutionary paths. Due to the limited sample size of SyXRBs, a deeper understanding of this sub-class requires further investigation.  

        \item The cumulative distribution functions (CDFs) of \vpec\, for BeXRBs and SgXRBs show clear differences, suggesting a preference for lower \vpec\, values in BeXRB systems compared to SgXRBs. A K-S test confirms that the two sub-groups are drawn from statistically distinct distributions.  

        \item For systems with neutron star accretors, \vpec-based classifications (\be: \vpec $ < 40$\kms; \sg: \vpec $> 40$\kms) are broadly consistent with the location of these classes in the Corbet diagram.

        \item A test of the directionality of peculiar motions shows that the \vpec\, vectors of HMXBs are consistent with isotropy. The ratio $\langle V_{\rm pec,3D}\rangle / \langle V_{\rm pec,2D}\rangle$ for our sample closely matches the theoretical value of $4/\pi$, enabling reliable estimation of 3D space velocities from observed 2D motions.  

        \item A plausible explanation for the kinematic segration is the differing nature of the progenitor systems for Sg vs. Be systems at the instant of supernova, with correspondingly different orbital velocities and ejecta masses (higher in both respects for the SgXRBs). 
        
        \item Simulations of XRB progenitor systems should be able to test the above scenario, and we utilise population synthesis tests to confirm that \sg\ systems generally form with shorter pre-SN orbital periods and higher fractional mass loss than \be\ systems, supporting the observational trends. 

        \item Irrespective of its physical cause, our empirical results imply that the magnitude of peculiar velocities could potentially be used as a complementary feature for identifying unclassified HMXBs.

        \item Our result represents 36 per cent of the confirmed HMXB population in our Galaxy. Despite potential selection effects due to parallax uncertainties and sample completeness, comparisons of absolute magnitude and colour distributions, supported by K-S tests, suggest that our sample of 63 systems is representative of the 172 known HMXBs.
        
    \end{enumerate}
    
\section*{Acknowledgements}
We thank the referee for their careful review and valuable feedback, which have significantly strengthened this paper. This work has made use of data from the European Space Agency (ESA) mission {\it Gaia} (\url{https://www.cosmos.esa.int/gaia}), processed by the {\it Gaia} Data Processing and Analysis Consortium (DPAC, \url{https://www.cosmos.esa.int/web/gaia/dpac/consortium}). Funding for the DPAC has been provided by national institutions, in particular the institutions participating in the {\it Gaia} Multilateral Agreement. PN is partially supported by funding from the Thai Government’s Development and Promotion of Science and Technology Talents Project (DPST), the National Astronomical Research Institute of Thailand (NARIT) and the University of Southampton. She also thanks Malcolm Coe and Phil Charles for the discussion during the initial stages. PG is a Royal Society Senior Leverhulme Trust fellow ($\backslash$R1$\backslash$241074), and also thanks STFC (ST/Y001680/1) for support. 

\section*{Data Availability}
The \gaia\ astrometry data can be accessed by querying the Gaia archive. The calculated \vpec\ values will be made available upon reasonable request to the corresponding author.




\bibliographystyle{mnras}
\bibliography{references} 




\section*{Appendix}

\begin{figure*}
	\includegraphics[width=0.8\textwidth]{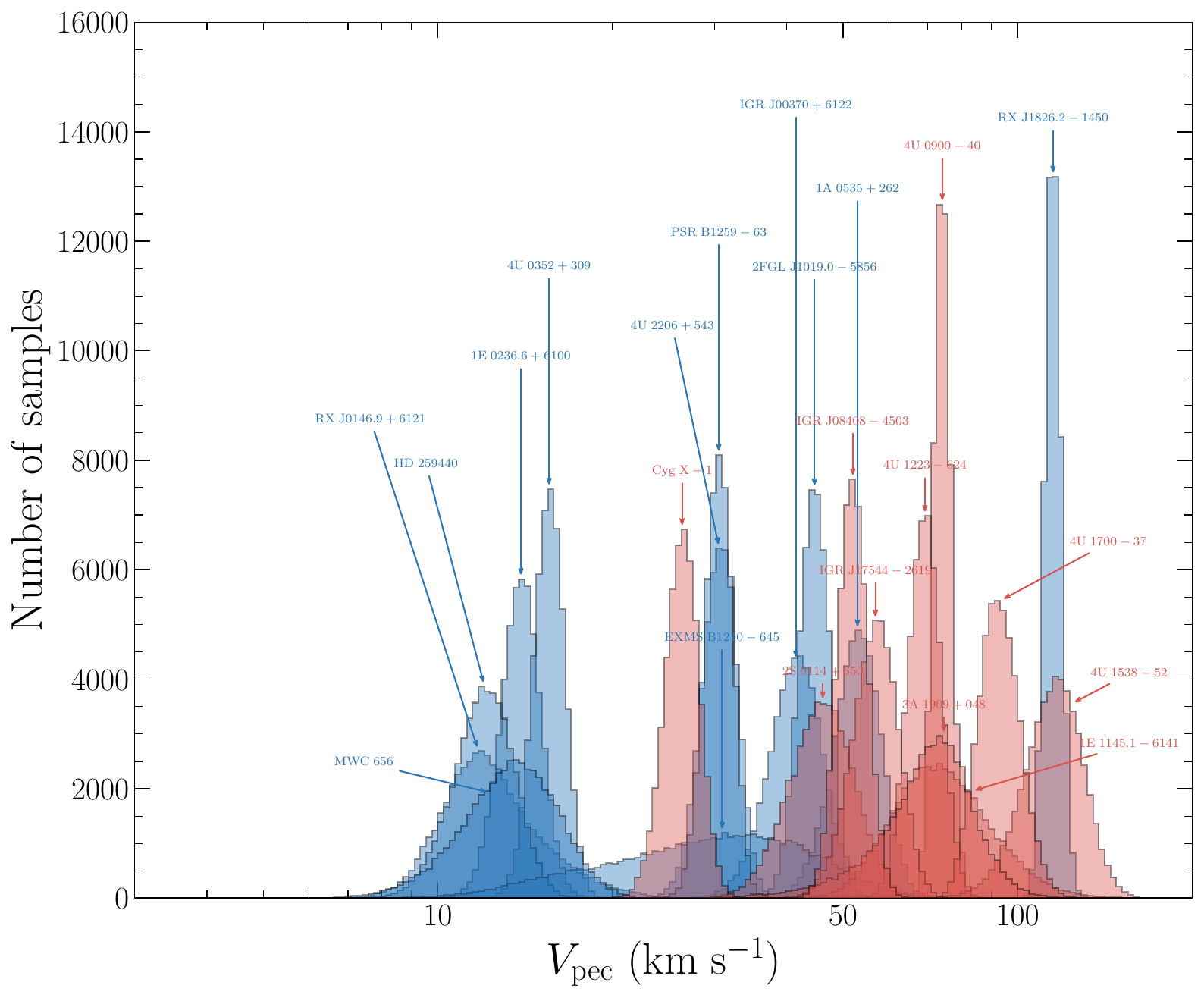}
    \caption{Individual posterior distributions of $V_{\rm pec}$ for BeXRBs (blue) and SgXRBs (red). Only systems with available $V_{\rm r}$ values are included here. For each source, 50,000 random samples are drawn.}
    \label{vp_2class_dis_identity}
\end{figure*}

\begin{figure*}
	\includegraphics[width=0.8\textwidth]{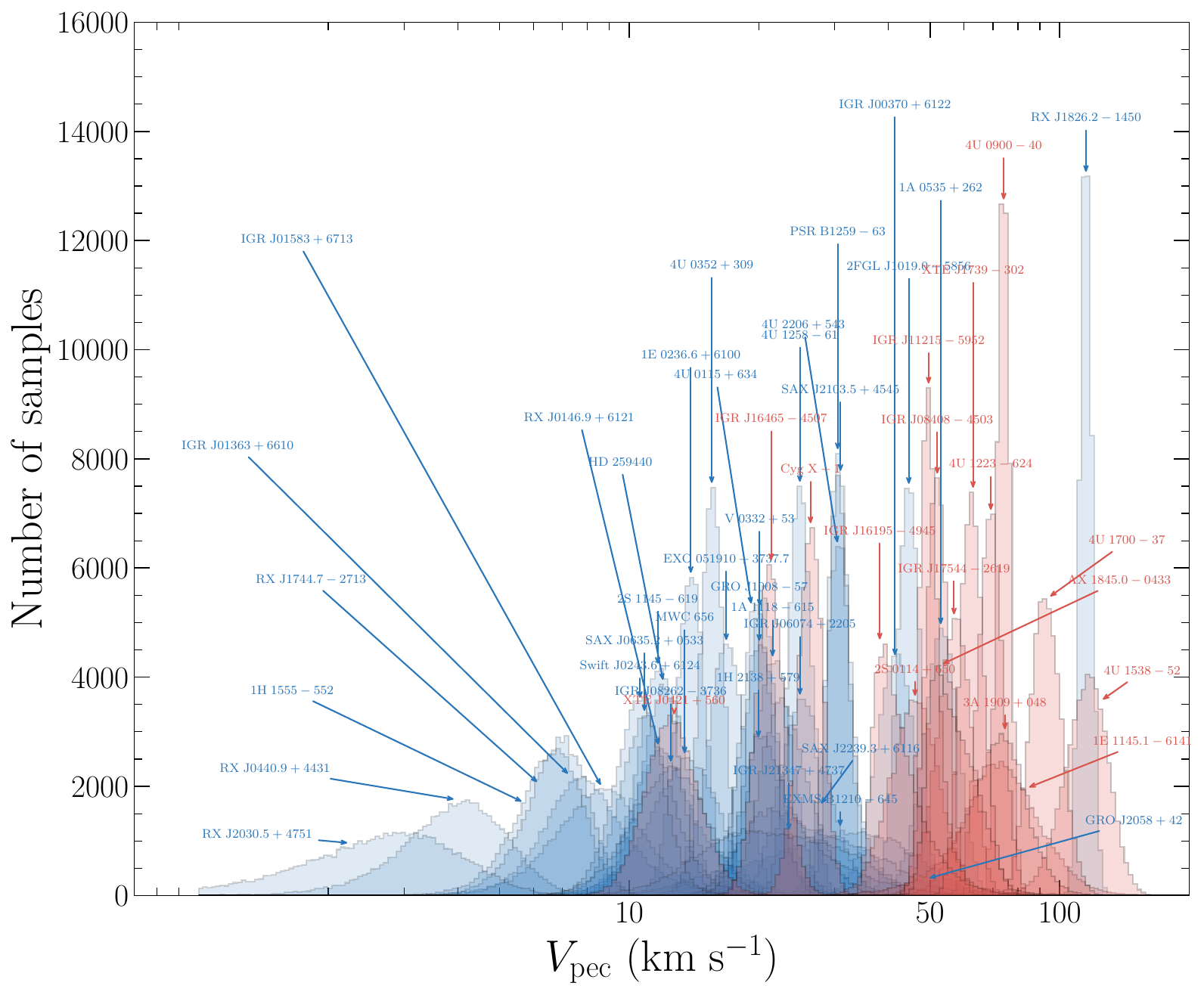}
    \caption{Individual probability distributions of $V_{\rm pec}$ for BeXRBs (blue) and SgXRBs (red). Systems with assumed $V_{\rm r}$ values using the isotropic assumption are included here. For each source, 50,000 random samples were drawn for plotting the probability distribution.}
    \label{vp_2class_dis_inc_unk_identity}
\end{figure*}

\begin{figure*}
	\includegraphics[width=0.8\textwidth]{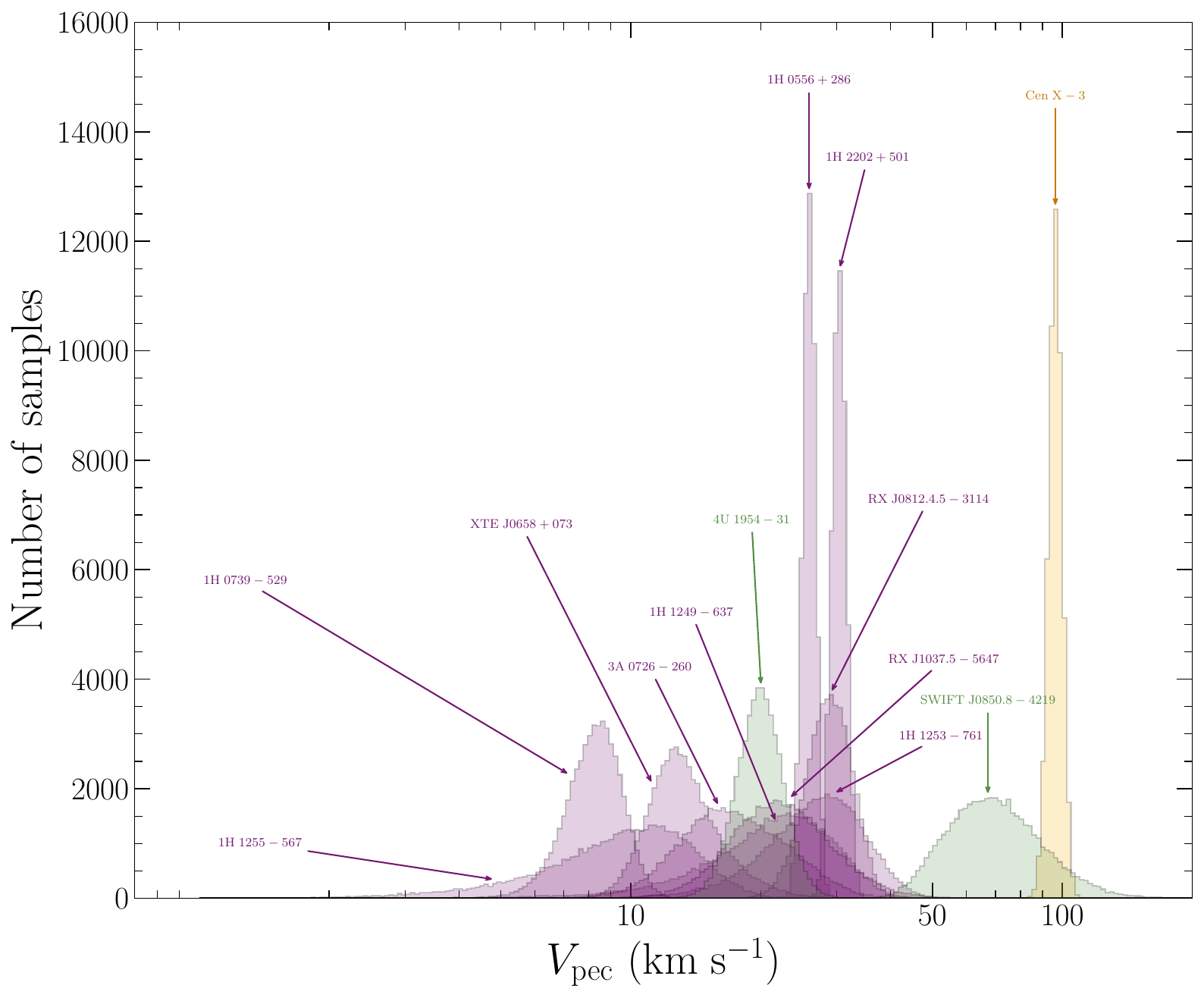}
    \caption{Individual probability distributions of $V_{\rm pec}$ for RLOs (yellow), SyXRBs (green) and unclear classes (margenta). Systems with assumed $V_{\rm r}$ values using isotropic assumption are included here. For each source, 50,000 random samples are drawn.}
    \label{vp_2class_dis_inc_unk_identity-rlo_sy_unc}
\end{figure*}

\subsection{Additional discussion of specific HMXBs}
\subsubsection{BH systems}
\label{sec:bhdiscussion}

Throughout our analyses, we include 4 systems that may host a BH, and some of their classification is still under some debate. In this section, we present individual discussion on these systems.\\

\noindent
\underline{RX J1826.2--1450/LS 5039}: This system is a HMXB first discovered by \citet{motch_new_1997}. Its optical counterpart, LS 5039, is an O star \citep[6.5V((f)), ][]{clark_radio_2001} and is in a 4.4-day orbit with a compact companion  \citep{mcswain_orbit_2001, mcswain_n_2004}. \citet{casares_possible_2005} reported the mass of the compact object to be $3.7_{-1.0}^{+1.3}$ $M_{\sun}$, suggesting that the compact object as a BH. Recent studies found signs of X-ray pulsations, so the compact object is more likely a NS \citep{yoneda_sign_2020, volkov_nustar_2021}. \citet{fortin_constraints_2022} found the \vpec\, of $89.1_{-2.6}^{+2.8}$ \kms; our calculation yields a consistent \vpec\ of 90.3$_{-2.9}^{+3.0}$ \kms, making it the fastest BeXRB in our sample. 
Furthermore, a recent study by \citet{Zeng_2024} suggests that LS 5039 may be a triple system, with a third body orbiting the barycentre of the binary. Gravitational interactions, including oscillations or dynamical ejection during close encounters, could alter the system’s orbit and explain its high \vpec. However, triple systems are not expected to significantly contribute to the high-velocity regime, as such systems typically require tight orbits, whereas the configuration of this system likely involves a third body in a wide orbit.\\

\noindent
\underline{3A 1909+048/SS 433}: This is a well-known HMXB hosting a supergiant and a compact object in an 13.1-day orbit \citep{crampton_ss_1981}. A compact object mass 2.9 $M_{\sun}$, which suggests that it might be a stellar-mass BH in the mass gap \citep{hillwig_identification_2004}. The nature of the compact object in SS 433 is still under debate. Spectroscopic observations estimating the compact object’s orbital speed, combined with the mass ratio from X-ray data, suggested that the compact object is likely a NS \citep{dodorico_evidence_1991}. Its \vpec\, is 57.0$\pm$10 \kms. It is often highlighted as the prototypical Galactic ultra-luminous X-ray source, plausibly a neutron star undergoing super-critical accretion, though this remains under debate  \citep[e.g., ][]{blundell_ss433, middleton_ss433}.\\

\noindent
\underline{4U 1956+35/Cyg X--1}: Cyg X--1 was the first BH HMXB to be identified in X-rays, and is now known to host a BH of approximately 21.2 $M_{\sun}$ and a O9.7 Iab supergiant with a mass of $\approx M_{\sun}$ \citep{miller-jones_cygnus_2021} in a close (5.6-day) orbit \citep{gies_wind_2003}. Its \vpec\, 20.8$_{-1.4}^{+1.4}$ \kms, which is quite low velocity for \sg. This could be attributed to the evolutionary path of Cyg X--1. BH in Cyg X--1 was form by implosion where system may not have experienced an energetic trigger from NK or significant mass loss associated with a SN event \citep{Mirabel_and_Rodriguez_2003}. In this study, we use the same method applied to other systems to calculate the \vpec\, of Cyg X--1 relative to the Galactic centre. Since Cyg X--1 is associated with the massive star cluster Cygnus OB3 \citep{Mirabel_and_Rodriguez_2003}, which is considered its parent association, the peculiar velocity should ideally be measured relative to Cygnus OB3. Previous studies report a \vpec\ of $\approx$\,9$\pm$2 \kms\ relative to Cygnus OB3 \citep{Mirabel_and_Rodriguez_2003, rao20}. Thus, the \vpec\, we derive here, being relative to the Galactic centre, can be considered as an upper limit.\\

\noindent
\underline{MWC 656/HD 215227}: This is a BeXRB with an orbital period of 60.37$\pm$0.04 days \citep{williams_be_2010, paredes-fortuny_optical_2012, casares_binary_2012, casares_be-type_2014}. Studies of the optical counterpart and spectral type of the secondary suggested a distance of 2.6$\pm$0.6 kpc \citep{casares_be-type_2014} and also indicated that the compact object in the system is a BH with a mass of 3.8--6.9 $M_{\sun}$, making MWC 656 the first known Be/BH system \citep{casares_be-type_2014}. 
However, \citet{rivinius_mwc656_2022} revisited the spectral variability properties of MWC 656 and concluded that it is more likely to be a hot subdwarf rather than a BH. This conclusion was further supported by \citet{Janssens_2023}, that the compact object in this system is not a black hole from spectroscopic data with high-Resolution Mercator Echelle Spectrograph (\textit{HERMES}). Similar to other BeXRBs in our sample, MWC 565 has a low \vpec\, (23.9$\pm$10 \kms). Such a low velocity suggests that the BH may have formed through direct collapse, without experiencing a NK from a SN explosion, similar to the case of Cyg X--1.
\\

\subsubsection{SyXRBs}
We include two rare cases of symbiotic HMXBs in our sample, both with no available $V_{\rm r}$ measurements. Our calculation reveals substantial different \vpec s for these two systems.\\

\noindent
\underline{4U 1954+31}: This system was discovered by the \emph{Uhuru}\, \emph{(SAS A)}\, mission \citep{Forman1978}. The early X-ray position has substantial uncertainty, which encloses multiple counterparts, including a Be star \citep{Tweedy1989}. A significantly more precise position was reported by \emph{Chandra}\, observations, identifying this system with an M-type star \citep{Masetti_2006_Mstar}, the spectral type was also confirmed by its near-infrared spectrum \citep{Hinkle2020}. The estimated mass of the donor star is approximately 9$_{-2}^{+6}$ $M_{\odot}$ \citep{Hinkle2020}. We derive a low \vpec\, of 19.7$_{-2.1}^{+2.2}$ \kms, consistent with a mild kick received at the instant of a supernova. This result supports the assumption of \citet{Hinkle2020}, suggesting that the supernova might ablate the surface of the B-type main-sequence companion. Alternatively, the subsequent mixing of surface material into the envelope could have occurred, causing the B-type main-sequence star to evolve into an M supergiant.\\

\noindent
\underline{Swift J0850.8--4219/2MASS J08504008--4211514}: This system was recently discovered by {\it Swift/XRT} as the second Galactic SyXRB. 
A possible near-infrared counterpart, 2MASS 08504008--4211514, corresponds to an red supergiant (RSG) of spectral type K3-K5 with an estimated distance of $\approx$12 kpc \citep{De_2023}. This system has a high \vpec\, of 68.8$\pm$19 \kms. If the companion turns out to be of low mass, this could be explained with a small system inertia, making it less resistant to acceleration by a natal kick and consequently leading to a relatively higher peculiar velocity. But this remains to be tested, as the system is still relatively new and ill-understood at the time of writing.

\subsubsection{Promising systems}
\label{sec:J0243iscussion}
\underline{Swift J0243.6+6124}: This system was recently discovered by {\it Swift/BAT} as the first and, to date, only ULXP identified within our Galaxy \citep{Kennea2017ATel,Doroshenko2018,Tsygankov2018,Wilson2018}. Its pulsation period of approximately 9.86~s has been confirmed by observations from {\it Swift/XRT} \citep{Kennea2017ATel}, {\it Fermi/GBM} \citep{JenkeWilson2017ATel}, and {\it NuSTAR} \citep{Bahramian2017ATel}. Optical spectroscopy initially identified the source as a new \be\ \citep{Kouroubatzakis2017}, with subsequent analysis classifying the optical companion as an O9.5Ve star \citep{J0243Reig2020}. Photometric measurements of the optical counterpart suggest a distance of $4.5\pm0.5$~kpc \citep{J0243Reig2020}.

There is only a single published $V_{\rm r}$ measurement available from SDSS/APOGEE, reporting a notably high value of 325.71 \kms\ \citep{Jonsson2020}, which subsequently yields a very large \vpec\ estimate of 393.76 \kms\ as reported by \citet{Wang2025}. Given the lack of corroborating $V_{\rm r}$ measurements, the reliability of this single measurement is uncertain. Therefore, we conservatively choose not to adopt this $V_{\rm r}$ value. Instead, we employ our ansatz approach of computing an isotropic estimate of \vpec, resulting in a substantially lower $V_{\rm pec,3D}^{\rm iso}$ of 10.4 \kms. With this assumption, the source comfortably aligns with the BeXRB sub-group, consistent with the typical velocities for other in BeXRB sub-group.


\bsp	
\label{lastpage}
\end{document}